\documentclass[aps,prl,amssymb,superscriptaddress,twocolumn]{revtex4-1}
\usepackage[pdftex]{graphicx}
\usepackage[dvipsnames]{xcolor}
\usepackage{bm}
\usepackage{amsmath}
\usepackage{ascmac}
\usepackage{color}
\usepackage{setspace}
\usepackage{amssymb}
\usepackage{latexsym}
\usepackage{mathtools}
\usepackage{lipsum}
\usepackage{braket}
\usepackage[colorlinks,linkcolor=PineGreen,citecolor=PineGreen,urlcolor=PineGreen,]{hyperref}

\begin{document}

\title{Impact of Dissipation on Universal Fluctuation Dynamics in Open Quantum Systems } 

\author{Kazuya Fujimoto}
\affiliation{Institute for Advanced Research, Nagoya University, Nagoya 464-8601, Japan}
\affiliation{Department of Applied Physics, Nagoya University, Nagoya 464-8603, Japan}

\author{Ryusuke Hamazaki}
\affiliation{Nonequilibrium Quantum Statistical Mechanics RIKEN Hakubi Research Team, RIKEN Cluster for Pioneering Research (CPR), RIKEN iTHEMS, Wako, Saitama 351-0198, Japan}

\author{Yuki Kawaguchi}
\affiliation{Department of Applied Physics, Nagoya University, Nagoya 464-8603, Japan}

\date{\today}

\begin{abstract}
Recent experimental and theoretical works have uncovered nontrivial quantum dynamics due to external dissipation. Using an exact numerical method and a renormalization-group-based analytical technique, we theoretically elucidate that dissipation drastically alters universal particle-number-fluctuation dynamics related to surface-roughness growth in non-interacting fermions and bosons. In a system under dephasing that causes loss of spatial coherence, we find that a universality class of surface-roughness dynamics changes from the ballistic class to a class with the Edwards-Wilkinson scaling exponents and an unconventional scaling function. On the other hand, in a system under dissipation with in- and out-flow of particles that breaks particle-number conservation, the universal dynamics is lost. 
\end{abstract}

\maketitle

{\it Introduction.}
Interactions with environments inevitably induce dissipation, giving rise to drastic changes in a quantum system \cite{open_quantum1,open_quantum2,open_quantum3,open_quantum4}. 
Such open quantum systems have attracted many theoretical and experimental researchers in recent years, and nontrivial consequences of dissipation have been uncovered in a variety of quantum phenomena. In particular, universal dynamics has been extensively explored from various viewpoints such as measurement-induced phase transitions \cite{measurement1,measurement2,measurement3,measurement4,measurement5,measurement6,measurement7,measurement8,measurement9,measurement10,measurement11,measurement12}, self-organized criticality \cite{SOC1,SOC2,SOC3}, and non-equilibrium phase transitions \cite{NPT1,NPT2,NPT21,NPT3,NPT4,NPT5,NPT6, NPT7, NPT8, NPT9, NPT10, NPT11, NPT12, NPT13,NPT14,NPT15,NPT16,NPT17,NPT18}. Open quantum systems have thus become a novel playground for studying universal quantum non-equilibrium phenomena. 
 
\begin{figure}[t]
\begin{center}
\includegraphics[keepaspectratio, width=8.6cm]{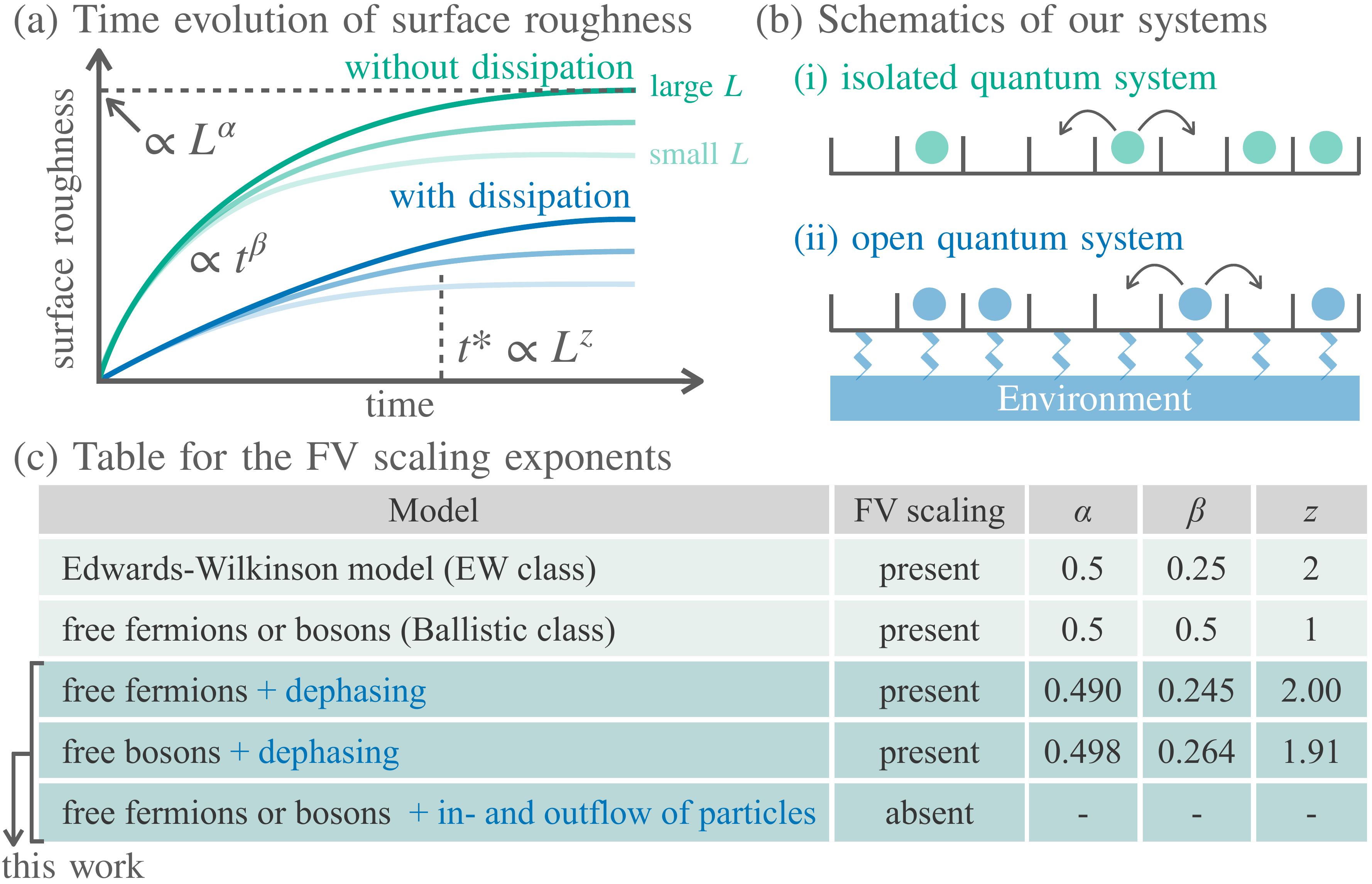}
\caption{
(a) Schematic illustration for the Family-Vicsek (FV) scaling with the scaling exponents $\alpha$, $\beta$, and $z$. The abscissa is time, and the ordinate is the surface roughness $w(L,t)$, which is a function of a system size $L$ and time $t$. The curves with the same color are the surface roughness with different system sizes. The roughness grows with $t^{\beta}$ for $t \ll t^*$, where the saturation time $t^*$ scales as $L^{z}$. The saturated surface roughness is proportional to $L^{\alpha}$.  In an open system, the roughness grows slowly compared with an isolated system. (b) Schematic illustrations for (i) an isolated system and (ii) an open system. In this work, we consider open systems with on-site dissipation. (c) Table for the presence of the FV scaling and its scaling exponents. The first two rows are the previously known universality classes, while the remaining three rows are the main results in this Letter. We here show the exponents for a staggered initial state. The values of the exponents obtained in this work are almost independent of initial states (see Table~\ref{table:exponents}). 
} 
\label{figure1} 
\end{center}
\end{figure}
 
One of the important universal dynamical phenomena, which is well known in classical open systems, is surface growth \cite{barabasi1995fractal, tauber2014}. A minimal theoretical model for the classical surface growth is the Kardar-Parisi-Zhang (KPZ) equation \cite{OKPZ}, which exhibits universal dynamical scaling in the surface-height distribution. Recent works found a signature of KPZ universality even in isolated quantum many-body systems, by investigating two-point spatio-temporal correlation functions numerically~\cite{Spin_KPZ1,Spin_KPZ2,Spin_KPZ3,Spin_KPZ4,Spin_KPZ5,Spin_KPZ6, Spin_KPZ7, Spin_KPZ8} and experimentally~\cite{Spin_KPZ_exp1,Spin_KPZ_exp2}. Instead of computing the correlation function, a surface-height operator and quantum surface roughness were introduced in Refs.~\cite{height_op1,height_op2,height_op3} by using the particle-number and spin fluctuations. Considering the particle-number fluctuations of isolated fermionic and bosonic lattice models, our previous works \cite{height_op1,height_op3} found emergence of the Family-Vicsek (FV) scaling \cite{Vicsek1984,Vicsek1985}, the dynamical scaling of the surface roughness originally developed in classical surface growth \cite{barabasi1995fractal}. As illustrated in Fig.~\ref{figure1}(a), this scaling is characterized by three exponents $\alpha$, $\beta$, and $z$, which determine universality classes of the dynamics \cite{barabasi1995fractal}.

In this Letter, we theoretically tackle a fundamental and intriguing question: ``How does dissipation affect the universal fluctuation dynamics related to the surface-growth physics in quantum systems?'' We consider an open quantum system with on-site dissipation as depicted in Fig.~\ref{figure1}(b), which obeys the Lindblad equation. To overcome the difficulty of investigating large-scale long-time dynamics by directly solving the Lindblad equation, we use an exact numerical method with correlation matrices \cite{NPT10,correlation_matrix1} and a renormalization-group-based analytical technique \cite{renormalization1,renormalization2,renormalization3}. First, studying non-interacting fermions and bosons on a one-dimensional (1D) lattice under dephasing, we numerically find that the FV scaling emerges even in the open quantum system and that the dissipation changes the universality class from the ballistic class to a class with the Edwards-Wilkinson (EW) scaling exponents (see Fig.~\ref{figure1}(c)) and an unconventional scaling function. We derive effective equations via the renormalization-group method and analytically explain the change of the universal scaling exponents. To the best our knowledge, this is the first analytical example to explain the FV scaling exponent in quantum systems. Second, we consider the Lindblad equation that breaks the particle-number conservation and numerically demonstrate the absence of the FV scaling. We conjecture that this is attributed to the absence of the slow dynamics induced by the particle-number conservation. Figure~\ref{figure1}(c) summarizes our results. Finally, we discuss experimental possibilities for observing our theoretical predictions. 

{\it Setup.}
We consider non-interacting fermions or bosons on a 1D lattice $\Lambda=\{1,2,\cdots,L\}$ with an even number $L$ of the lattice points. 
Let $\hat{a}_j$ and $\hat{a}_j^{\dagger}$ be the annihilation and creation operators at a site $j \in \Lambda$. When the particles are fermions (bosons), the operators satisfy $ [ \hat{a}_j, \hat{a}_k^{\dagger} ]_{+} = \delta_{jk}$ ($ [ \hat{a}_j, \hat{a}_k^{\dagger} ]_{-} = \delta_{jk}$), where we introduce the (anti)commutator $[ \hat{A}, \hat{B} ]_{\pm} = \hat{A} \hat{B} \pm \hat{B} \hat{A}$ for operators $\hat{A}$ and $\hat{B}$. 
We assume that a quantum state at time $t$, specified by a density matrix $\hat{\rho}(t)$, obeys the Lindblad equation \cite{open_quantum1}:
\begin{eqnarray}
\frac{d}{dt} \hat{\rho}(t) = -i[\hat{H}, \hat{\rho}(t)]_{-} + \mathcal{D}[\hat{\rho}(t)], 
\label{EOM}
\end{eqnarray}
where $\hat{H}$ and $\mathcal{D}[\hat{\rho}(t)]$ are respectively a Hamiltonian and a dissipator.  
The Hamiltonian $\hat{H}$, which describes coherent dynamics for the non-interacting particles, is given by
$\hat{H} = -  \sum_{j=1}^{L-1} \left( \hat{a}_{j+1}^{\dagger} \hat{a}_{j} + \hat{a}_{j}^{\dagger} \hat{a}_{j+1} \right)$. 
In this work, we consider two kinds of on-site dissipators. 
One is the dephasing dissipator conserving the total particle number, which is defined by
\begin{eqnarray}
\mathcal{D}_{\rm dep} [\hat{\rho}(t)] =  \gamma \sum_{j=1}^{L} \left(   \hat{n}_j  \hat{\rho}(t) \hat{n}_j - \frac{1}{2}[ \hat{n}_j^2, \hat{\rho}(t) ]_{+} \right) 
\label{dissipator1}
\end{eqnarray}
with a particle-number operator $\hat{n}_j = \hat{a}^{\dagger}_j \hat{a}_j$ and a strength $\gamma$ of the dephasing. 
The other is the dissipator describing in- and out-flow of the particles:
\begin{eqnarray}
\mathcal{D}_{\rm in/out} [\hat{\rho}(t)] = 
&&  \gamma_{\rm in} \sum_{j=1}^{L} \left(   \hat{a}_j^{\dagger}  \hat{\rho}(t) \hat{a}_j - \frac{1}{2}[ \hat{a}_j \hat{a}_j^{\dagger}, \hat{\rho}(t) ]_{+} \right) \nonumber \\
&& + \gamma_{\rm out} \sum_{j=1}^{L} \left(   \hat{a}_j  \hat{\rho}(t) \hat{a}_j ^{\dagger} - \frac{1}{2}[ \hat{a}_j^{\dagger} \hat{a}_j, \hat{\rho}(t) ]_{+} \right), 
\label{dissipator2}
\end{eqnarray}
where $\gamma_{\rm in}$ and $\gamma_{\rm out}$ are parameters for the in- and out-flow of particles. Using this model, we study the dynamics starting from a staggered state (SS) $\ket{{\rm SS}} = \prod_{j=1}^{L/2} \hat{a}^{\dagger}_{2j} \ket{0}$, a domain-wall state (DWS) $\ket{{\rm DWS}} = \prod_{j=1}^{L/2} \hat{a}^{\dagger}_{j} \ket{0}$, and a uniform state (US) $\ket{{\rm US}} = \prod_{j=1}^{L} \hat{a}^{\dagger}_{j} \ket{0}$ with the vacuum $\ket{0}$. 

\begin{figure*}[t]
\begin{center}
\includegraphics[keepaspectratio, width=18cm]{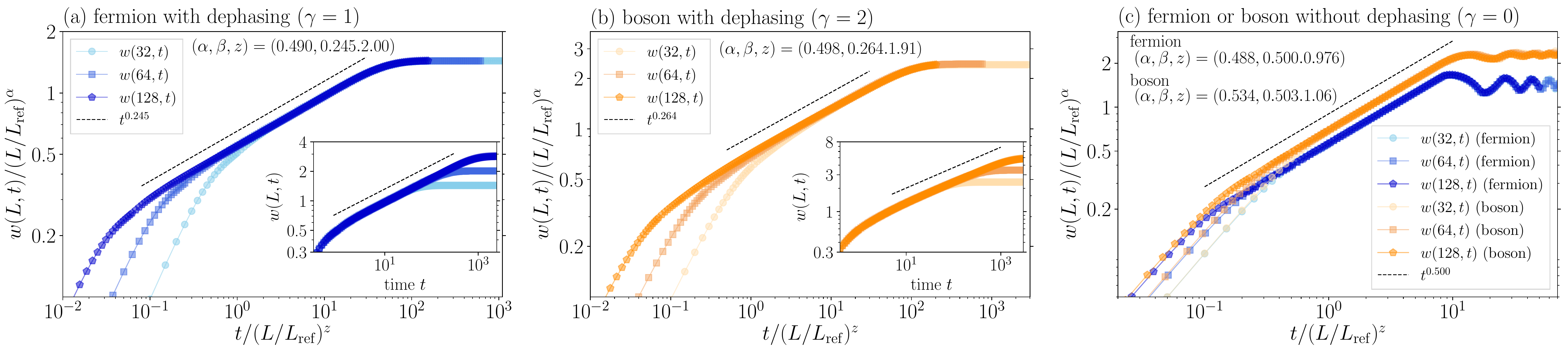}
\caption{Time evolution of the surface roughness for (a) fermions with $\gamma=1$, (b) bosons with $\gamma=2$, and (c) fermions and bosons with $\gamma=0$. In the main panel, we show the surface roughness with the ordinate and abscissa normalized by $(L/L_{\rm ref})^{\alpha}$ and $(L/L_{\rm ref})^{z}$ with $L_{\rm ref}=32$. The estimated exponents $(\alpha,\beta,z)$ are (a) (0.490, 0.245, 2.00), (b) (0.498, 0.264, 1.91), and (c) (0.488, 0.500, 0.976) for the fermion and (0.534, 0.503, 1.06) for the boson.
} 
\label{figure2} 
\end{center}
\end{figure*}

The physical quantity of interest is surface roughness defined by a variance of a surface-height operator $\hat{h}_j = \sum_{k=1}^{j} \left( \hat{n}_k - \nu \right)$ with an initial filling factor $\nu$ \cite{height_op1,height_op2,height_op3}. This operator was introduced on the basis of a mathematical correspondence between a classical surface height $h(x,t)$ in the KPZ equation and a sound mode $\delta n (x,t)$ (or a heat mode) in the fluctuating hydrodynamics in 1D systems. Since  scaling functions have the same form for $h(x,t)$ and $\partial_x \delta n (x,t)$ in the stationary processes, one can define an effective surface height by $h_{\rm eff}(x,t) = \int_{0}^{x} \delta n (y,t) dy $ in the fluctuating hydrodynamics. The surface-height operator $\hat{h}_j$ is quantum extension of $h_{\rm eff}(x,t)$. Using this surface-height operator, we define the surface roughness at a site $j \in \Lambda$ by $w_j(t) = \sqrt{ \braket{ ( \hat{h}_j - \braket{\hat{h}_j}_t )^2 }_t }$ with the quantum statistical average $\braket{\cdots}_{t} = {\rm Tr}[\hat{\rho}(t) \cdots]$. In what follows, we focus on $j=L/2$ because the surface roughness grows for longer time for this choice than the other $j$ and introduce the notation $w(L,t) = w_{L/2}(t)$ for brevity.

When the fluctuation of the surface height is scale-invariant, the surface roughness shows FV scaling \cite{Vicsek1984,Vicsek1985,barabasi1995fractal} defined by 
\begin{eqnarray}
w(L,t) = s^{-\alpha} w(sL, s^{z}t) \propto 
\left\{
\begin{array}{ll}
t^{\beta}   & (t \ll t^{*}) \\
L^{\alpha} & (t^{*} \ll t).
\end{array}
\right.
\end{eqnarray}
Here, $\alpha$, $\beta$, and $z$ are scaling exponents classifying a universality class, and $t^*$ is saturation time. Two well-known universality classes originally found in classical systems are the KPZ \cite{OKPZ} and EW classes \cite{OEW} characterized by $(\alpha,\beta,z) = (1/2,1/3,3/2)$ and $(1/2,1/4, 2)$, which show superdiffusive ($1<z<2$) and diffusive ($z=2$) transport, respectively. In quantum systems, non-interacting fermions have $(\alpha,\beta,z) \simeq (1/2,1/2,1)$ \cite{height_op1,height_op3}, which we call a ballistic class since the dynamical exponent $z$ is unity.

{\it Numerical method.}
Instead of directly solving the Lindblad equation, we solve the equations of motion for two- and four-point correlation matrices \cite{NPT10,correlation_matrix1} defined by $D_{mn} = \braket{\hat{a}_m^{\dagger} \hat{a}_n}_t$ and $F_{mnpq} = \braket{\hat{a}_m^{\dagger} \hat{a}_n^{\dagger} \hat{a}_p \hat{a}_q}_t$, respectively. 
As shown in Sec.~I of Supplemental Material (SM)~\cite{SM}, we exactly derive the closed equations of motion, which enable us to access the  long-time universal dynamics in the open quantum systems. 
Note that a third quantization and a superoperator method are other well-known efficient techniques to solve the Lindblad equation \cite{lindblad_ex1,lindblad_ex2,lindblad_ex21,lindblad_ex3,lindblad_ex4,lindblad_ex5,lindblad_ex6,lindblad_ex7}, but they are inconvenient in the dehaping case since this dissipation generates quartic terms in the thermofield representation.

Solving the equations for the correlation matrices, we can calculate the surface roughness using the following formula:
\begin{eqnarray}
w(L,t)^2 = && \pm \sum_{m=1}^{L/2} \sum_{n=1}^{L/2} F_{mnmn}(t) + (1- \nu L) \sum_{m=1}^{L/2} D_{mm}(t) \nonumber \\
&& + \frac{ \nu^2 L^2}{4} - \left( \sum_{m=1}^{L/2}  D_{mm}(t) -  \frac{\nu L}{2} \right)^2. 
\label{roughness_formula}
\end{eqnarray}
Here, $-(+)$ sign is for fermions (bosons). 

{\it Results of the Lindblad equation for the dephasing.}
We study how the dephasing~\eqref{dissipator1} affects the surface-roughness dynamics. In this model, the dynamics occurs in a sector with a fixed total particle number, and hence, we expect that the fluctuation of the local particle number exhibits slow dynamics whose time scale increases with the system size $L$.

Figure~\ref{figure2} shows time evolution of the surface roughness for (a) fermions with $\gamma=1$, (b) bosons with $\gamma=2$, and (c) fermions and bosons with $\gamma=0$. The initial state is SS. From Figs.~\ref{figure2}(a) and (b), one can see that the FV scaling is well satisfied for both fermions and bosons with the dephasing, and the scaling exponents are almost the same as those for the EW class. Interestingly, we find that the scaling function has an unconventional form being different from the EW equation as discussed in Sec.~II of SM~\cite{SM}. On the other hand, the isolated fermions and bosons show the FV scaling with the ballistic class~\cite{height_op1,height_op3} as shown in Fig.~\ref{figure2}(c). Our numerical results clearly show that the dephasing alters the universality class from the ballistic class to the one with the EW-type exponents characterized by the diffusive dynamics with $z=2$. 

\begin{figure}[b]
\begin{center}
\includegraphics[keepaspectratio, width=8.5cm,clip]{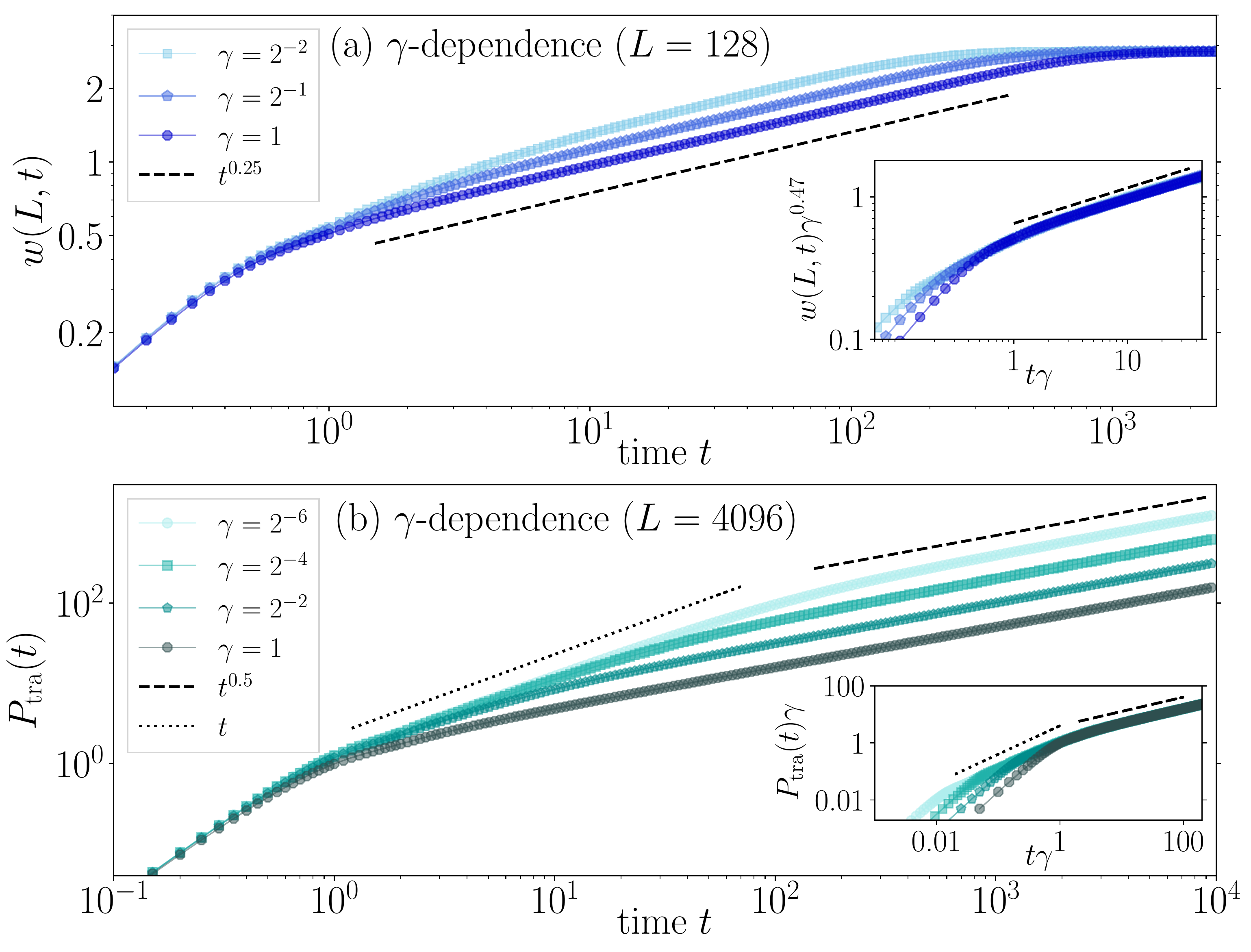}
\caption{Dependence of the fermionic dynamics on $\gamma$. (a) Time evolution of $w(128,t)$ for $\gamma=1, 2^{-1},$ and $2^{-2}$. The initial state is SS. In the inset, we normalize the time and the roughness by $1/\gamma$ and $1/\gamma^{0.47}$. This clearly exhibits that $\beta=0.25$, the signature of the EW class (diffusive dynamics),  emerges in $t \gtrsim 1/\gamma$. (b) Time evolution of $P_{\rm tra}(t)$ for $\gamma=1, 2^{-2}, 2^{-4},$ and $2^{-6}$. The initial state is DWS, and the system size is $4096$. In the inset, we normalize the time and the roughness by $1/\gamma$ and $1/\gamma$. Similar to (a), the diffusive behavior ($P_{\rm tra}(t) \propto t^{0.5}$) emerges for $t \gtrsim 1/\gamma$. } 
\label{figure3} 
\end{center}
\end{figure}

Next, we investigate dependence of the dynamics on $\gamma$. Figure~\ref{figure3}(a) shows the time evolution of $w(L,t)$ for fermions with $\gamma=1,2^{-1}$, and $2^{-2}$. We find that the surface roughness obeys $t^{0.25}$ in the late dynamics ($t \gtrsim 1/\gamma$), which corresponds to the EW exponent. This fact is clearly seen in the inset of Fig.~\ref{figure3}(a). From this result, we argue that the change of the universality class occurs for infinitesimal dissipation strength $\gamma$, which indicates the strong impact of dissipation. To strengthen our argument, we also study whether the diffusive transport emerges in the dynamics starting from the DWS. As discussed in Ref.~\cite{Spin_KPZ_exp2}, the particle transfer from the left to right region is used to study the transport property. Here, we numerically compute $P_{\rm tra}(t) = (N_{\rm R}(t) - N_{\rm R}(0)) - (N_{\rm L}(t)-N_{\rm L}(0))$ with $N_{\rm L}(t) = \sum_{m=1}^{L/2} \braket{\hat{a}^{\dagger}_m \hat{a}_m }_t$ and $N_{\rm R}(t) = \sum_{m=L/2+1}^{L} \braket{\hat{a}^{\dagger}_m \hat{a}_m }_t$.  If the transport is diffusive, we have $P_{\rm tra}(t) \propto t^{0.5}$. Figure~\ref{figure3}(b) shows the time evolution of $P_{\rm tra}(t)$ for the fermions with $\gamma=1,2^{-2},2^{-4}$, and $2^{-6}$ in the large system size $L=4096$. This result demonstrates that, for $\gamma \gtrsim 2^{-6}$, the ballistic behavior appears in the early dynamics ($1/\gamma \gtrsim t \gtrsim 1$) but the transport eventually becomes diffusive for sufficiently long time ($t \gtrsim 1/\gamma$). This numerical finding strongly supports our argument. 

\begin{table}[b]
  \caption{
  FV scaling exponents for the dephasing model of fermions ($\gamma=1$) and bosons ($\gamma=2$). The obtained FV scaling exponents are close to the exponents of the EW class. The fitting error are 3$\sigma$ error evaluated in the method of Ref.~\cite{expoents_cal}. 
  }
  \label{table:exponents}
  \centering
  \begin{tabular}{lccc}
    \hline
    Initial state  & $\alpha$  &  $\beta$ & $z$  \\
    \hline \hline
    [fermion] \\
    SS   &  0.490 $\pm$ 0.035  & 0.245 $\pm$ 0.005 & 2.00 $\pm$ 0.15 \\
    DWS  &   0.483 $\pm$ 0.061  & 0.249 $\pm$ 0.006 & 1.96 $\pm$ 0.26 \\
    \hline \hline
    [boson] \\
    SS     & 0.498 $\pm$ 0.067  & 0.264 $\pm$ 0.007 & 1.91 $\pm$ 0.29 \\
    DWS   & 0.503 $\pm$ 0.036  & 0.260 $\pm$ 0.005 & 1.95 $\pm$ 0.16 \\
    US        & 0.499 $\pm$ 0.052  & 0.262 $\pm$ 0.006 & 1.92 $\pm$ 0.23 \\
    \hline
  \end{tabular}
\end{table}

We numerically investigate the dependence of the FV scaling exponents on the initial states. The detailed time-evolution of the surface roughness is given in Sec.~III of SM~\cite{SM}. The obtained exponents are summarized in Table~\ref{table:exponents}, which shows that the exponents are almost independent of the initial states if the initial surface roughness is small. 

To understand the change of the universality class analytically, we use a perturbative renormalization-group method \cite{renormalization1,renormalization2,renormalization3,renormalization4} and derive effective equations for $ D_{mm}$ and $F_{mnmn}$, which determine the surface roughness~\eqref{roughness_formula}. As derived in Sec.~IV of SM~\cite{SM}, when the dephasing strength $\gamma$ is strong, the effective equations for the fermions and the bosons become
\begin{eqnarray}
&& \frac{d}{dt} D_{mm}  \simeq \frac{2}{\gamma} (  {D}_{(m+1)(m+1)} + {D}_{(m-1)(m-1)}  -  2 {D}_{mm} ),  \\ 
&& \frac{d}{dt} F_{mnmn}  \simeq \frac{2}{\gamma} ( F_{(m+1)n(m+1)n}  + F_{(m-1)n(m-1)n}  \nonumber \\
&&~~~~~~~~~~~~~ + F_{m(n+1)m(n+1)} + F_{m(n-1)m(n-1)}  -  4 F_{mnmn} )  \nonumber \\
\label{f_F11}
\end{eqnarray}
for $|m-n| > 2$. Taking the continuum limit for these equations, we obtain the one- and two-dimensional diffusion equations, which are responsible for the diffusive transport ($z=2$). As discussed in Sec.~IV of SM~\cite{SM}, the effective equations show the similar dynamics to the exact numerical results. Our renormalization-group analysis explains the emergence of universal scaling behavior with the EW exponents. As far as we know, this is the first example to explain the FV scaling exponents in quantum systems analytically.

While the above method is only for strong $\gamma$, in Sec.~V of SM~\cite{SM}, making several assumptions and focusing on the late stage of the dynamics ($t \gtrsim 1/ \gamma$), we derive the same effective equations without assuming the strong dephasing. This supports the emergence of the EW scaling exponents for infinitesimally small dephasing.

{\it Results of the Lindblad equation for the in- and out-flow of particles.}
We next discuss how the breaking of the particle-number conservation affects the surface-roughness dynamics by considering the in- and out-flow type of dissipation \eqref{dissipator2}. In this case, denoting the total-particle-number operator by $\hat{N}_{\rm tot}(t) = \sum_{m=1}^{L} \hat{a}_m^{\dagger} \hat{a}_m$, we can show that $\braket{ \hat{N}_{\rm tot} }_{t}$ and $\braket{ \hat{N}_{\rm tot}^2 }_{t}$ are generally dependent on time. For example, the fermionic system obeys $d_t \braket{ \hat{N}_{\rm tot} }_{t} = \gamma_{\rm in} L  - (\gamma_{\rm  out}+\gamma_{\rm  in}) \braket{ \hat{N}_{\rm tot} }_{t} $ while the bosonic system obeys $d_t \braket{ \hat{N}_{\rm tot} }_{t}  = \gamma_{\rm in} L  - (\gamma_{\rm  out} - \gamma_{\rm  in}) \braket{ \hat{N}_{\rm tot} }_{t} $. 

\begin{figure}[t]
\begin{center}
\includegraphics[keepaspectratio, width=8.5cm,clip]{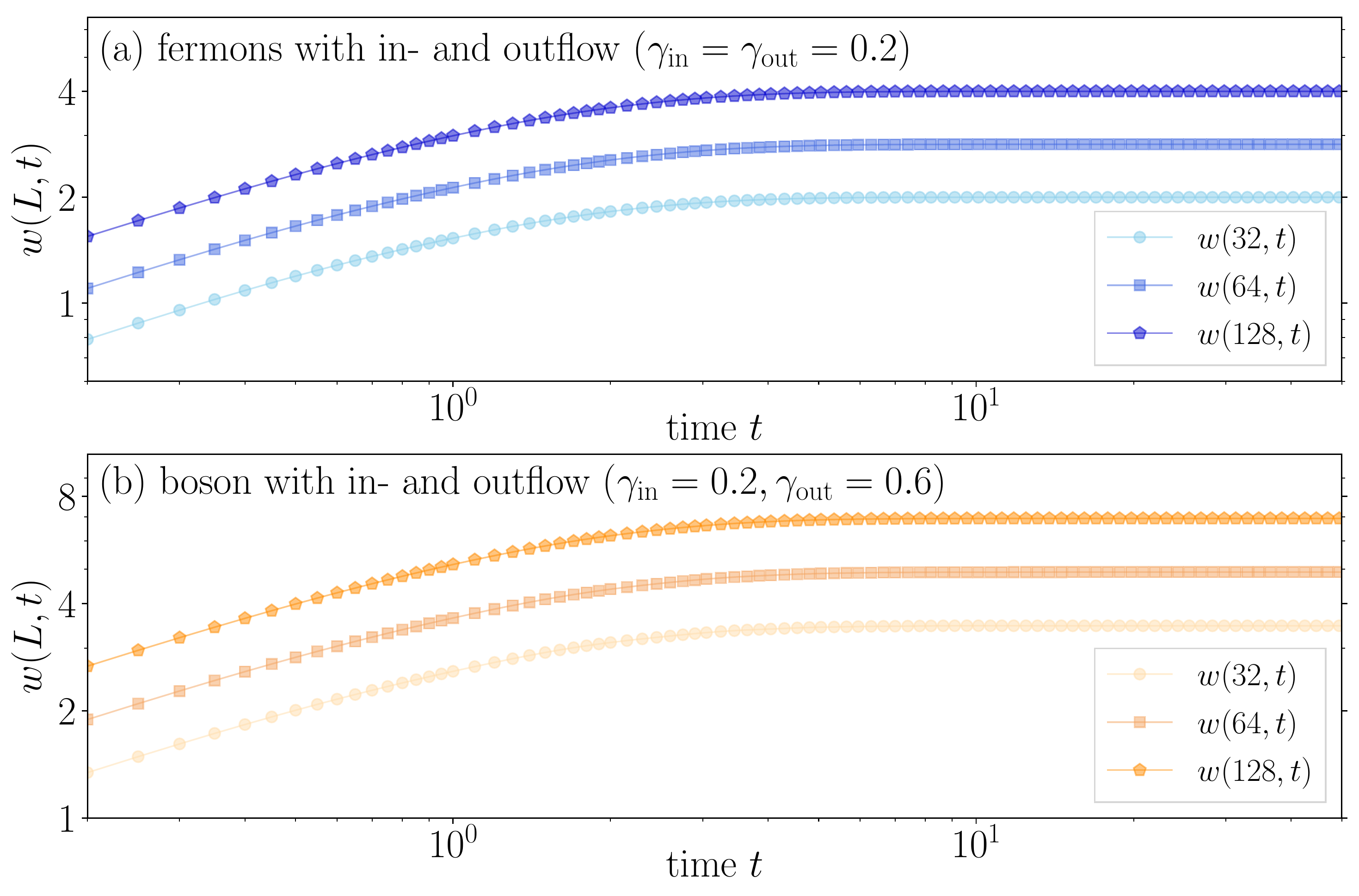}
\caption{
Surface-roughness dynamics in fermions and bosons with the in- and out-flow of particles for $L=32, 64$, and $128$. The initial state is SS. (a) Fermionic dynamics with $\gamma_{\rm in} = \gamma_{\rm out} = 0.2$. (b) Bosonic dynamics with $\gamma_{\rm in} = 0.2$ and $\gamma_{\rm out} = 0.6$. As far as we investigate, the FV scaling is not found even when we use other choices of $\gamma_{\rm in}$ and  $\gamma_{\rm out}$. 
} 
\label{figure4} 
\end{center}
\end{figure}

Figure~\ref{figure4} shows the time evolution of the surface roughness with the in- and out-flow of particles. The initial state is the SS, and the parameters are $\gamma_{\rm in}=\gamma_{\rm out} = 0.2$ for the fermions and $ \gamma_{\rm in} = 0.2$ and $\gamma_{\rm out}=0.6$ for the bosons, which are chosen such that $\braket{ \hat{N}_{\rm tot} }_{t}$ is conserved in time but $\braket{ \hat{N}_{\rm tot}^2 }_{t}$ is not. Figure~\ref{figure4} exhibits that the saturation time of the surface roughness is independent of the system size. This finding clearly shows that the Lindblad equation for the in- and out-flow of particles does not exhibit the conventional FV scaling associated with large-scale fluctuation growth characterized by nonzero dynamical exponent $z$.

We conjecture that the absence of the FV scaling is attributed to the breaking of the particle-number conservation. Conservation laws often lead to slow dynamics and the appearance of  physical quantities whose timescale grows with $L$. The absence of the conservation laws can break this slow dynamics and the FV scaling, as demonstrated in our numerical simulations.  

{\it Discussion.}
We discuss experimental possibilities for observing our theoretical predictions. An experiment under photon scattering \cite{Dephasing_exp1} is considered to be well described by the Lindblad equation for the dephasing~\eqref{dissipator1}. Thus, the change of universality class can be experimentally accessible when one observes dynamics beyond the dephasing timescale $1/\gamma$. As to the in- and out-flow of particles~\eqref{dissipator2}, the particle loss can be experimentally controlled by illuminating cold atoms \cite{Loss1,Loss2}, and thus the disappearance of the FV scaling, which is discussed in Sec.~VI of SM~\cite{SM}, may be detectable.

{\it Concluding remarks and future prospects.}
We theoretically studied the surface-roughness dynamics described by the Lindblad equation with the two types of dissipation: one is the dephasing, and the other is the in- and out-flow of particles. In the dephasing case, we numerically found the emergence of a clear FV scaling with a novel scaling function and analytically elucidated that the universality class is altered due to the presence of the dephasing. From these numerical and analytical results, we argued that the change of the universality class occurs at infinitesimally small dissipation, suggesting the substantial impact of dissipation. In the in- and out-flow case, the FV scaling does not emerge, which we conjectured is caused by the breaking of the particle-number conservation. 

Our findings pave an intriguing avenue for exploring the universal fluctuation dynamics in open quantum systems. In this work, we found the big impact of the dissipation on the universal surface-roughness dynamics in the fundamental non-interacting models; it is interesting to study whether open quantum many-body systems exhibit novel universal dynamics triggered by the interactions. It is also important to consider other dissipation such as incoherent hopping \cite{transport1,transport2, transport3,transport4,transport5,transport6}, which have close relation to classical stochastic processes, e.g., an asymmetric simple exclusion process \cite{ASEP1,ASEP2}.

\begin{acknowledgments}
We are grateful to Hosho Katsura for helpful comments on the manuscript.
This work was supported by JST-CREST (Grant No. JPMJCR16F2), JSPS KAKENHI (Grant Nos. JP18K03538, JP19H01824,  JP19K14628, 20H01843, 21H01009), the Toyota Riken Scholar program, Foundation of Kinoshita Memorial Enterprise, and the Program for Fostering Researchers for the Next Generation (IAR, Nagoya University) and Building of Consortia for the Development of Human Resources in Science and Technology (MEXT). 
\end{acknowledgments}

\bibliography{reference}

\widetext
\clearpage

\setcounter{equation}{0}
\setcounter{figure}{0}
\setcounter{section}{0}
\setcounter{table}{0}
\renewcommand{\theequation}{S-\arabic{equation}}
\renewcommand{\thefigure}{S-\arabic{figure}}
\renewcommand{\thetable}{S-\arabic{table}}

\section*{Supplemental Material for ``Impact of Dissipation on Universal Fluctuation Dynamics in Open Quantum Systems''}
This Supplemental Material describes the following topics:
\begin{itemize}
\item[  ]{ (I) Numerical method, } 
\item[  ]{ (II) Scaling function of the Edwards-Wilkinson equation, }
\item[  ]{ (III) Numerical data for the growing surface roughness and its dependence on the initial states, }
\item[  ]{ (IV) Renormalization-group derivation of the effective diffusive equations for the dephasing Lindblad equation, }
\item[  ]{ (V) Derivation of the effective diffusive equations for the Lindblad equation without the strong-dephasing condition, }
\item[  ]{ (VI) Surface-roughness dynamics only with the outflow of particles. }
\end{itemize}

\section{Numerical method}
We describe how to numerically solve the Lindblad equation for the density matrix $\hat{\rho}(t)$ at a time $t$. A basic idea is to derive equations of motion for the two- and four-point correlation matrices defined by
\begin{eqnarray}
D_{mn}(t) := \braket{ \hat{a}_{m}^{\dagger} \hat{a}_{n} }_t, 
\end{eqnarray}
\begin{eqnarray}
F_{mnpq}(t) := \braket{ \hat{a}_{m}^{\dagger} \hat{a}_{n}^{\dagger} \hat{a}_{p} \hat{a}_{q} }_t. 
\end{eqnarray}
Here, the bracket means the average $\braket{\cdots}_t = {\rm{Tr}}[\hat{\rho}(t) \cdots]$.

\subsection{Lindblad equation with the dephasing}
Using the Lindblad equation with the dephasing in the main text, we can derive the equations of motion for $D_{mn}(t)$ and $F_{mnpq}(t)$ \cite{NPT10,correlation_matrix1}. For an operator $\hat{A}$ comprised of the fermionic or bosonic operators, we obtain 
 \begin{eqnarray}
\frac{d}{d t} \braket{ \hat{A} }_t =  -i \braket{ [\hat{A},\hat{H}]_- }_t + \frac{\gamma}{2} \sum_{j=1}^{L} \left\langle \left[ \hat{n}_j, \left[\hat{A},\hat{n}_j \right]_-  \right]_-  \right\rangle_t, 
\label{A_eq1}
\end{eqnarray}
where we use the fact that the jump operator is Hermitian.
One can see that the equations of motion for $D_{mn}(t)$ and $F_{mnpq}(t)$ can be closed if $\hat{H}$ is the Hamiltonian used in the main text. 

\subsubsection{fermionic system}
When the annihilation and creation operators follow the fermionic anticommutation relation, the equation~\eqref{A_eq1} becomes
\begin{eqnarray}
\frac{d}{d t} D_{mn} = i\left( D_{m(n+1)} + D_{m(n-1)} - D_{(m+1)n}  - D_{(m-1)n}  \right) + \gamma \left( \delta_{mn} - 1 \right) D_{mn}, 
\label{f_D}
\end{eqnarray}
\begin{eqnarray}
\frac{d}{d t} F_{mnpq} =~&& i (  F_{mn(p+1)q} + F_{mn(p-1)q} + F_{mnp(q+1)} + F_{mnp(q-1)}  - F_{(m+1)npq} - F_{(m-1)npq} - F_{m(n+1)pq} - F_{m(n-1)pq} )  \nonumber \\
&& + \gamma \left( \delta_{mq} + \delta_{mp}  + \delta_{nq} + \delta_{np}  - 2   \right) F_{mnpq}.
\label{f_F}
\end{eqnarray}

\subsubsection{bosonic system}
When the annihilation and creation operators follow the bosonic commutation relation, the equation~\eqref{A_eq1} becomes
\begin{eqnarray}
\frac{d}{d t} D_{mn} = i\left( D_{m(n+1)} + D_{m(n-1)} - D_{(m+1)n}  - D_{(m-1)n}  \right) + \gamma \left( \delta_{mn} - 1 \right) D_{mn}, 
\label{b_D}
\end{eqnarray}
\begin{eqnarray}
\frac{d}{d t} F_{mnpq} =~&& i (  F_{mn(p+1)q} + F_{mn(p-1)q} + F_{mnp(q+1)} + F_{mnp(q-1)}  - F_{(m+1)npq} - F_{(m-1)npq} - F_{m(n+1)pq} - F_{m(n-1)pq} )  \nonumber \\
&& + \gamma \left( \delta_{mq} + \delta_{np} + \delta_{mq} + \delta_{nq} -  \delta_{mn}  - \delta_{pq}   - 2   \right) F_{mnpq}.
\label{b_F}
\end{eqnarray}

\subsection{Lindblad equation for the in- and out-flow of particles}
Using the Lindblad equation for the in- and out-flow of particles in the main text, we can derive the equations of motion for $D_{mn}(t)$ and $F_{mnpq}(t)$.
For an operator $\hat{A}$ comprised of the fermion or boson operators, we obtain 
 \begin{eqnarray}
\frac{d}{d t} \braket{ \hat{A} }_t =  -i \braket{ [\hat{A},\hat{H}]_- }_t + \gamma_{\rm in} \sum_{j=1}^{L} \left\langle  \hat{a}_j \hat{A} \hat{a}^{\dagger}_j - \frac{1}{2}[ \hat{a}_j \hat{a}^{\dagger}_j, \hat{A} ]_+   \right\rangle_t + \gamma_{\rm out} \sum_{j=1}^{L} \left\langle  \hat{a}^{\dagger}_j \hat{A} \hat{a}_j - \frac{1}{2}[ \hat{a}_j^{\dagger} \hat{a}_j, \hat{A} ]_+   \right\rangle_t.
\label{A_eq2}
\end{eqnarray}
This equation has a different form from Eq.~\eqref{A_eq1} because the jump operators for the in- and out-flow of particles are not Hermitian. 

\subsubsection{fermionic system}
The straightforward calculations with Eq.~\eqref{A_eq2} lead to 
\begin{eqnarray}
\frac{d}{d t} D_{mn} = i\left( D_{m(n+1)} + D_{m(n-1)} - D_{(m+1)n}  - D_{(m-1)n}  \right) + \gamma_{\rm in} \left( \delta_{mn} - D_{mn} \right) - \gamma_{\rm out} D_{mn}, 
\end{eqnarray}
\begin{eqnarray}
\frac{d}{d t} F_{mnpq} =~&& i (  F_{mn(p+1)q} + F_{mn(p-1)q} + F_{mnp(q+1)} + F_{mnp(q-1)}  - F_{(m+1)npq} - F_{(m-1)npq} - F_{m(n+1)pq} - F_{m(n-1)pq} )  \nonumber \\
&& + \gamma_{\rm in} \left( \delta_{mq}D_{np} + \delta_{np}D_{mq} - \delta_{nq}D_{mp} - \delta_{mp}D_{nq} - 2 F_{mnpq}  \right) - 2 \gamma_{\rm out} F_{mnpq}.
\end{eqnarray}

\subsubsection{bosonic system}
The straightforward calculations with Eq.~\eqref{A_eq2} lead to
\begin{eqnarray}
\frac{d}{d t} D_{mn} = i\left( D_{m(n+1)} + D_{m(n-1)} - D_{(m+1)n}  - D_{(m-1)n}  \right) + \gamma_{\rm in} \left( \delta_{mn} + D_{mn} \right) - \gamma_{\rm out} D_{mn}, 
\end{eqnarray}
\begin{eqnarray}
\frac{d}{d t} F_{mnpq} =~&& i (  F_{mn(p+1)q} + F_{mn(p-1)q} + F_{mnp(q+1)} + F_{mnp(q-1)}  - F_{(m+1)npq} - F_{(m-1)npq} - F_{m(n+1)pq} - F_{m(n-1)pq} )  \nonumber \\
&& + \gamma_{\rm in} \left(   2F_{mnpq}   + \delta_{mq}D_{np} + \delta_{nq} D_{mp} + \delta_{np} D_{mq} + \delta_{mp} D_{nq}  \right) - 2 \gamma_{\rm out} F_{mnpq}.
\end{eqnarray}
\clearpage

\section{Scaling function of the Edwards-Wilkinson equation}
In the main text, we show that the Lindblad equation for the dephasing well obeys the Family-Vicsek (FV) scaling with the Edwards-Wilkinson (EW) exponents. Here, we discuss the scaling function of the quantum surface-roughness by comparing it with a scaling function of the EW equation. 

\begin{figure}[b]
\begin{center}
\includegraphics[keepaspectratio, width=18.0cm,clip]{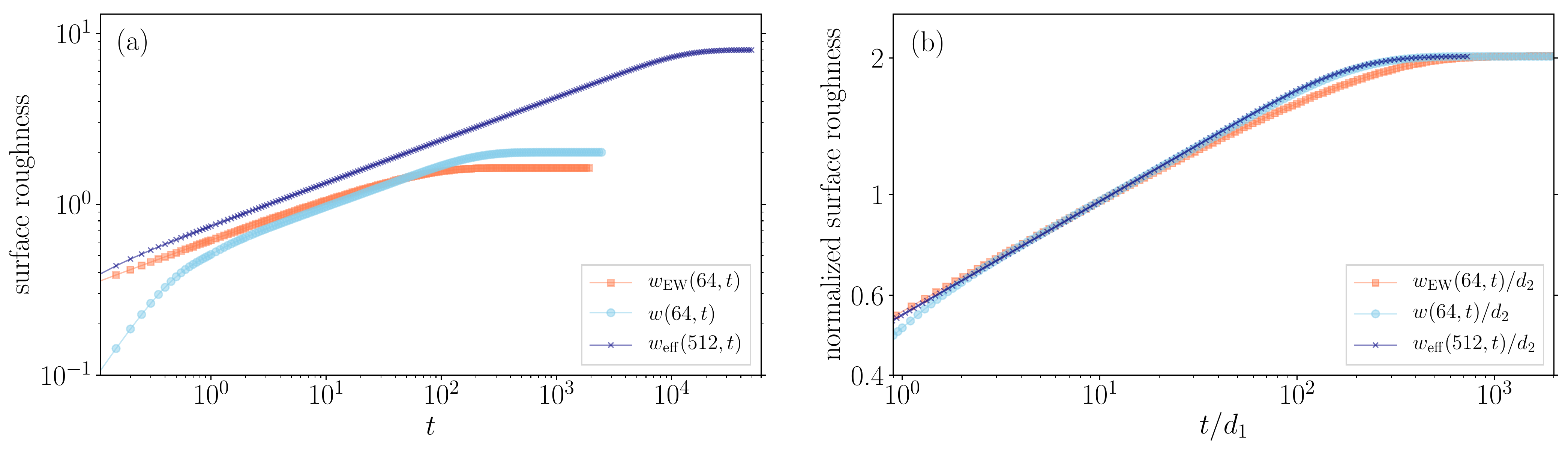}
\caption{
Time evolution of $w_{\rm EW}(L'=64,t)$ compared with the quantum surface roughness $w(L=64,t)$ and $w_{\rm eff}(L=512,t)$, which are obtained by solving the fermionic Lindblad equation for the dephasing ($\gamma=1$) and its effective equations~\eqref{sec_reno_perturbed_D6} and \eqref{sec_reno_perturbed_F14}, respectively. The initial state is the staggered state. (a) Raw data for $w_{\rm EW}(64,t)$, $w(64,t)$, and $w_{\rm eff}(512,t)$. (b) Comparison of $w_{\rm EW}(64,t)$ with $w(64,t)$ and $w_{\rm eff}(512,t)$. We normalize the abscissa and the ordinate by $d_1=1.00, 0.270, 68.0$ and $d_2=1.00, 0.809, 3.97$ for $w(64,t)$, $w_{\rm EW}(64,t)$, and $w_{\rm eff}(512,t)$. These normalization parameters are determined by fitting the three curves by eye. One can see that the scaling function of the EW equation is evidently different from that of the open quantum systems. 
} 
\label{figure_scaling_function} 
\end{center}
\end{figure}

We first review the EW equation and its scaling function.  
The one-dimensional EW equation is an equation of motion for a one-dimensional surface height $h(x,t)$, which is given by
\begin{eqnarray}
\frac{\partial}{\partial t} h(x,t) = \frac{\partial^2}{\partial x^2} h(x,t) + \xi(x,t),  
\end{eqnarray}
where the white noise $\xi(x,t)$ satisfies $\braket{\xi(x,t)}_{\rm en} =0$ and $\braket{\xi(x,t)\xi(x',t')}_{\rm en} = \delta(x-x') \delta(t-t')$. Here  $\braket{\cdots}_{\rm en}$ means the ensemble average over classical noise. The boundary condition is assumed to be periodic. Solving the stochastic differential equation via the Fourier transformation, we obtain the following expression of the classical surface roughness $w_{\rm EW}(L',t)$:
\begin{eqnarray}
w_{\rm EW}(L',t)^2
&:=& \frac{1}{L'} \int_{0}^{L'} dx \left\langle  \left(  h(x,t) -  \frac{1}{L'}  \int_{0}^{L'} dy h(y,t) \right)^2 \right\rangle_{\rm en} \\
&=& \frac{L'}{4\pi^2} \sum_{n=1}^{\infty} \left( \frac{1}{n^2} - \frac{1}{n^2} {\rm exp}\left( -\frac{8\pi^2}{L^2}n^2 t \right) \right) \\
&=& \frac{L'}{24} - \frac{L'}{4 \pi^2} \sum_{n=1}^{\infty} \frac{1}{n^2} \exp \left( -\frac{8 \pi^2}{L'^2} n^2 t \right) \\
&=& \displaystyle \frac{L'}{24} - \frac{L'}{4 \pi^2} \sum_{n=1}^{\infty} \left( \frac{1}{n^2} - \int^{ 8 \pi^2 t / L'^2 }_{0} dx  \exp \left( -n^2 x \right)  \right) \\
&=&  \frac{L'}{4 \pi^2}  \int^{ 8 \pi^2 t / L'^2 }_{0} dx  \sum_{n=1}^{\infty} \exp \left( -n^2 x \right)  \\
&=& \frac{L'}{8 \pi^2}  \int^{ 8 \pi^2 t / L'^2 }_{0} dx \left( \vartheta_3 (0,i x^2/\pi) - 1 \right).
\label{EW_scaling_function}
\end{eqnarray}
Here, $L'$ is the one-dimensional system size, $\vartheta_3 (\nu,\tau)$ is the elliptic theta function, and the integer $n$ in the summation is a label of the discrete wavenumber in the Fourier transformation.

We compare the scaling function~\eqref{EW_scaling_function} of the EW equation with the quantum surface-roughness $w(L=64,t)$ in Fig.~\ref{figure_scaling_function}, which is obtained by the fermionic Lindblad equation with the dephasing ($\gamma=1$) starting from the staggered initial state. This result clearly demonstrates that the scaling function of the open quantum system is different from that of the EW equation. In the large system size $L=512$ which is beyond the exact calculation, we numerically solve the effective equations~\eqref{sec_reno_perturbed_D6} and \eqref{sec_reno_perturbed_F14} for the Lindblad equation, which are analytically derived by the renormalization-group method in the following section~\ref{sec:RG_derivation}. From Fig.~\ref{figure_scaling_function}, we find that the effective surface roughness $w_{\rm eff}(L=512,t)$ has the scaling function almost identical to $w(L=64,t)$. All these results strongly support the conjecture that the scaling function of the dephasing Lindblad equation is different from that of the EW equation. Currently, we do not find the origin for this difference. 

\section{Numerical data for the growing surface roughness and its dependence on the initial states}
In Table I of the main text, we summarize the FV scaling exponents obtained by the dephasing Lindblad dynamics starting from a staggered state (SS), a domain-wall state (DWS), and a uniform state (US). Here, we show the detailed time evolution of the surface roughness not shown in the main text. 

\begin{figure}[b]
\begin{center}
\includegraphics[keepaspectratio, width=18cm,clip]{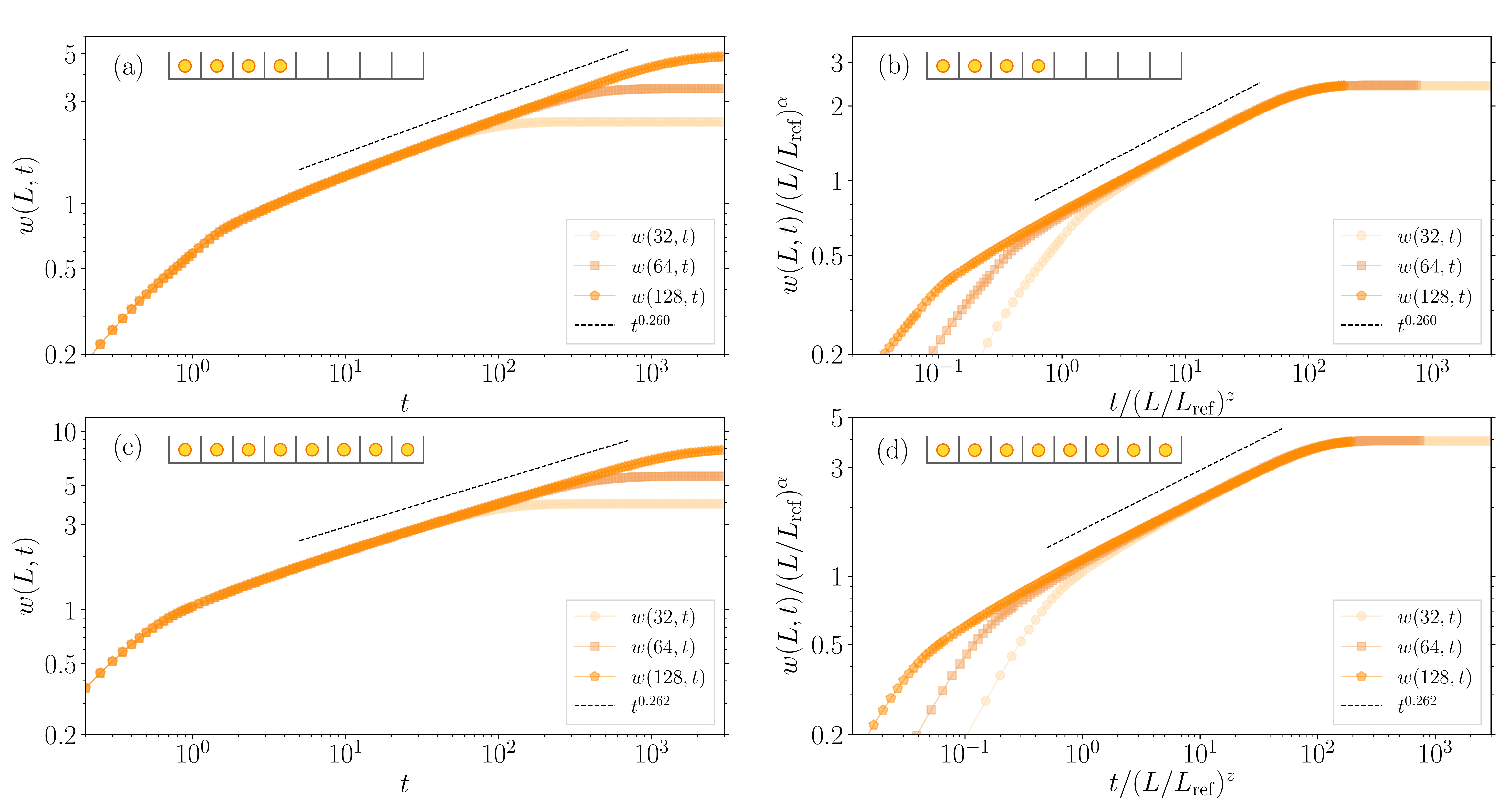}
\caption{
Surface-roughness dynamics in bosons with the dephasing ($\gamma=2$). The initial state used here is the domain-wall state for (a) and (b) and the uniform state for (c) and (d). Schematic illustrations of each initial state are depicted in the left upper region. The panels (a) and (c) are the raw data of the surface roughness for $L=32,64,$ and $128$. The panels (b) and (d) are the normalized figures for (a) and (c), respectively, where the ordinate and abscissa are normalized by $(L/L_{\rm ref} )^{\alpha}$ and $(L/L_{\rm ref} )^z$ with $L_{\rm ref} = 32$.  The evaluated exponents are $(\alpha,\beta,\gamma)=(0.503, 0.260, 1.95)$ for the domain-wall state and $(\alpha,\beta,\gamma)=(0.499, 0.262, 1.92)$ for the uniform state.
} 
\label{figureS4} 
\end{center}
\end{figure}

\begin{figure}[t]
\begin{center}
\includegraphics[keepaspectratio, width=18cm,clip]{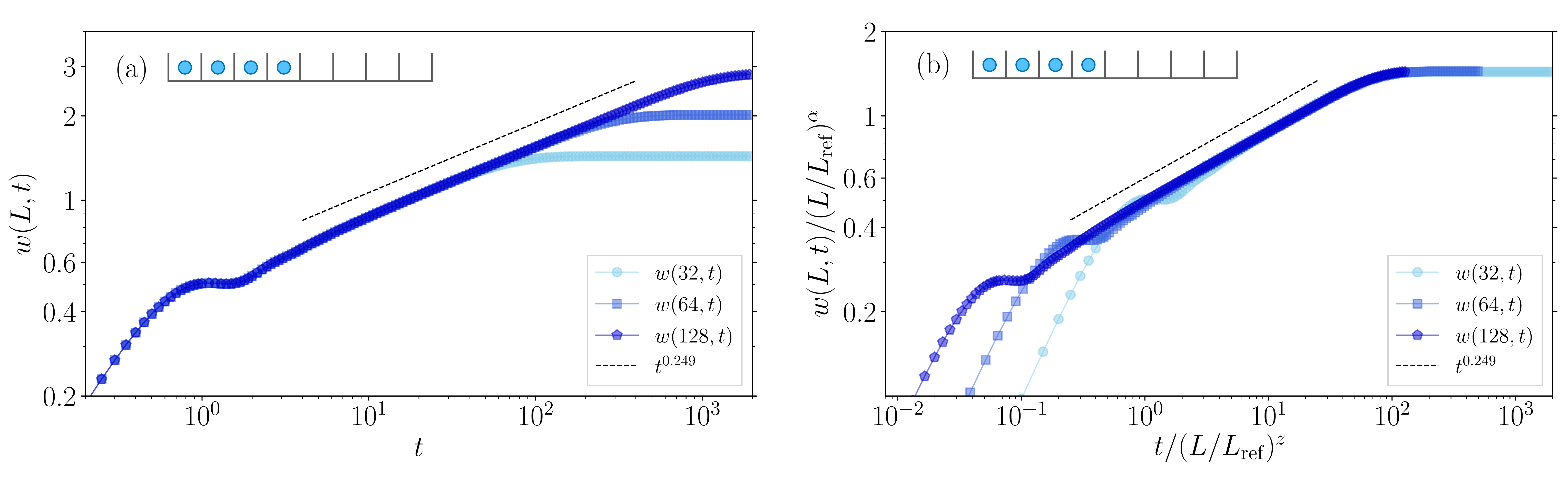}
\caption{Surface-roughness dynamics in fermions with the dephasing ($\gamma=1$). The initial state is the domain-wall state whose sketch is depicted in the left upper region. (a) Raw data of the surface roughness for $L=32,64,$ and $128$. (b) Family-Vicsek scaling for (a). The ordinate and abscissa are normalized by $(L/L_{\rm ref} )^{\alpha}$ and $(L/L_{\rm ref} )^z$ with $L_{\rm ref} = 32$. The evaluated exponents are $(\alpha,\beta,\gamma)=(0.483, 0.249, 1.96)$.
} 
\label{figureS3} 
\end{center}
\end{figure}

Figures~\ref{figureS4} and \ref{figureS3} display the time evolution in the bosons and fermions with the dephasing, respectively. For the bosonic system, Figs.~\ref{figureS4}(a,b) and (c,d) show the surface roughness for DWS and US. The strength of the dephasing is $\gamma=2$ and the system size is set to be $L=32, 64$, and $128$. Figures~\ref{figureS4}(a) and (c) are the raw numerical data, and Figs.~\ref{figureS4}(b) and (d) correspond to the figures with the normalized axes where we multiply the abscissa and the ordinate by $(L/L_{\rm ref})^{-z}$ and $(L/L_{\rm ref})^{-\alpha}$ with $L_{\rm ref}=32$. The evaluated FV scaling exponents are $(\alpha,\beta,\gamma)=(0.503, 0.260, 1.95)$ for DWS and $(\alpha,\beta,\gamma)=(0.499, 0.262, 1.92)$ for US. In the fermionic system, the initial state is DWS, and the parameters are set to be $\gamma=1$ and $L=32, 64$, and $128$. Figure~\ref{figureS3}(a) is the raw numerical data, and in Fig.~\ref{figureS3}(b) we similarly normalize the ordinate and abscissa to obtain the FV scaling. One can see the clear FV scaling with $(\alpha,\beta,\gamma)=(0.483, 0.249, 1.96)$.

\clearpage
\section{Renormalization-group derivation of the effective diffusive equations for the dephasing Lindblad equation}\label{sec:RG_derivation}
Employing a renormalization group method for a differential equation \cite{renormalization1,renormalization2,renormalization3}, we derive the effective equations of motion for the correlation matrices under the strong dephasing $\gamma \gg 1$. The physical quantity that we are interested in is the surface roughness, which is expressed by
\begin{eqnarray}
w(L,t)^2 = \pm \sum_{m=1}^{L/2} \sum_{n=1}^{L/2} F_{mnmn}(t) + (1- \nu L) \sum_{m=1}^{L/2} D_{mm} + \frac{\nu^2 L^2}{4} - \left( \sum_{m=1}^{L/2}  D_{mm}(t) -  \frac{\nu L}{2}  \right)^2. 
\end{eqnarray}
Here, $-(+)$ sign is the expression for fermions (bosons). 
One can see that obtaining $D_{mm}(t)$ and $F_{mnmn}(t)$ is sufficient to calculate the surface roughness. In the following, we derive effective equations of motion for $D_{mm}(t)$ and $F_{mnmn}(t)$. 

\subsubsection{fermionic system}
First, we derive the effective equation for $D_{mm}(t)$.  We introduce a small parameter $\epsilon = \gamma^{-1}$ and a new time variable $\tau = \gamma t$. Then, Eq.~\eqref{f_D} becomes
\begin{eqnarray}
\frac{d}{d \tau} \bar{D}_{mn}(\tau) =  \lambda_{mn} \bar{D}_{mn}(\tau) + i \epsilon \left( \bar{D}_{m(n+1)}(\tau) + \bar{D}_{m(n-1)}(\tau) - \bar{D}_{(m+1)n}(\tau)  - \bar{D}_{(m-1)n}(\tau)  \right) 
\label{sec_reno_f_D}
\end{eqnarray}
with $\bar{D}_{mn}(\tau) = D_{mn}(t)$ and $\lambda_{mn}= \delta_{mn} - 1 $. Following a conventional renormalization procedure, we expand the two-point correlation matrix as
\begin{eqnarray}
\bar{D}_{mn}(\tau) = \bar{D}_{mn}^{(0)}(\tau) + \epsilon \bar{D}_{mn}^{(1)}(\tau) + \epsilon^2 \bar{D}_{mn}^{(2)}(\tau) + \cdots, 
\label{sec_reno_expn1}
\end{eqnarray}
where $\bar{D}_{mn}^{(j)}(\tau)~(j=0,1,2,\cdots)$ is a $j$th-order two-point correlation matrix.
We substitute Eq.~\eqref{sec_reno_expn1} into \eqref{sec_reno_f_D} and obtain the following  equations:
\begin{eqnarray}
&&\epsilon^0:~~\frac{d}{d \tau} \bar{D}_{mn}^{(0)}(\tau) = \lambda_{mn} \bar{D}_{mn}^{(0)}(\tau),  \label{sec_reno_recursive_D1} \\
\nonumber \\
&&\epsilon^1:~~\frac{d}{d \tau} \bar{D}_{mn}^{(1)}(\tau) = \lambda_{mn} \bar{D}_{mn}^{(1)}(\tau) + i \left( \bar{D}_{m(n+1)}^{(0)}(\tau) + \bar{D}_{m(n-1)}^{(0)}(\tau) - \bar{D}_{(m+1)n}^{(0)}(\tau)  - \bar{D}_{(m-1)n}^{(0)}(\tau)  \right) , \label{sec_reno_recursive_D2} \\
\nonumber \\
&&\epsilon^2:~~\frac{d}{d \tau} \bar{D}_{mn}^{(2)}(\tau) = \lambda_{mn} \bar{D}_{mn}^{(2)}(\tau) + i \left( \bar{D}_{m(n+1)}^{(1)}(\tau) + \bar{D}_{m(n-1)}^{(1)}(\tau) - \bar{D}_{(m+1)n}^{(1)}(\tau)  - \bar{D}_{(m-1)n}^{(1)}(\tau)  \right), \label{sec_reno_recursive_D3} \\
&& ~\cdot \nonumber \\
&& ~\cdot \nonumber \\
&& ~\cdot \nonumber \\
&& \epsilon^a:~~\frac{d}{d \tau} \bar{D}_{mn}^{(a)}(\tau) = \lambda_{mn} \bar{D}_{mn}^{(a)}(\tau) + i \left( \bar{D}_{m(n+1)}^{(a-1)}(\tau) + \bar{D}_{m(n-1)}^{(a-1)}(\tau) - \bar{D}_{(m+1)n}^{(a-1)}(\tau)  - \bar{D}_{(m-1)n}^{(a-1)}(\tau)  \right). \label{sec_reno_recursive_D4}
\end{eqnarray}
A solution of Eq.~\eqref{sec_reno_recursive_D1} is given by
\begin{eqnarray}
\bar{D}_{mn}^{(0)}(\tau) = A_{mn} {\rm e}^{ \lambda_{mn} \tau }
\end{eqnarray}
with a constant $A_{mn}$. 
Using this solution, we solve the first-order equation~\eqref{sec_reno_recursive_D2}:
\begin{eqnarray}
\bar{D}_{mn}^{(1)}(\tau) =  i  {\rm e}^{ \lambda_{mn} \tau } \int_{0}^{\tau} dT {\rm e}^{ - \lambda_{mn} T } \left( \bar{D}_{m(n+1)}^{(0)}(T) + \bar{D}_{m(n-1)}^{(0)}(T) - \bar{D}_{(m+1)n}^{(0)}(T)  - \bar{D}_{(m-1)n}^{(0)}(T)  \right). 
\label{sec_reno_perturbed_D1}
\end{eqnarray}
When we put $n=m+1, m,  m-1, m + b$ with $|b|>1$, the above equation reads
\begin{eqnarray}
\bar{D}_{mm}^{(1)}(\tau) &=&  i \left( 1 - {\rm e}^{-\tau} \right)   \left( A_{m(m+1)} + A_{m(m-1)} - A_{(m+1)m} - A_{(m-1)m} \right), \label{first-order_D1} \\
\nonumber \\
\bar{D}_{m(m+1)}^{(1)}(\tau) &=&  i \tau {\rm e}^{-\tau}  \left( A_{m(m+2)} - A_{(m-1)(m+1)} \right) + i \left( 1 - {\rm e}^{-\tau} \right) \left( A_{mm} - A_{(m+1)(m+1)} \right), \label{first-order_D2} \\
\nonumber \\
\bar{D}_{m(m-1)}^{(1)}(\tau) &=&  i \tau {\rm e}^{-\tau}  \left( A_{m(m-2)} - A_{(m+1)(m-1)} \right) + i \left( 1 - {\rm e}^{-\tau} \right) \left( A_{mm} - A_{(m-1)(m-1)} \right), \label{first-order_D3}\\
\nonumber \\
\bar{D}_{m(m+b)}^{(1)}(\tau) &=&  i \tau {\rm e}^{-\tau}  \left( A_{m(m+b+1)} + A_{m(m+b-1)} - A_{(m+1)(m+b)} - A_{(m-1)(m+b)}  \right) \label{first-order_D4}. 
\end{eqnarray}
Up to the first order, all the perturbed solutions converge for $\tau \rightarrow \infty$.  However, in the second order, one can find divergent solutions. 
Substituting Eqs.~\eqref{first-order_D1}-\eqref{first-order_D4} into Eq.~\eqref{sec_reno_recursive_D3}, we obtain the second-order solution for the diagonal part:
\begin{eqnarray}
\bar{D}_{mm}^{(2)}(\tau) &=&  2 \tau \left( A_{(m+1)(m+1)} + A_{(m-1)(m-1)} - 2A_{mm} \right) + \left( {\rm convergent~term} \right), 
\label{sec_reno_perturbed_D2}
\end{eqnarray}
which diverges as $\tau$ goes to infinity. Then, we get the perturbed solution up to the second-order:
\begin{eqnarray}
\bar{D}_{mm}(\tau) = &&  A_{mm}  +  i \epsilon \left( 1 - {\rm e}^{-\tau} \right)   \left( A_{m(m+1)} + A_{m(m-1)} - A_{(m+1)m} - A_{(m-1)m} \right)\nonumber \\
&&+ 2 \epsilon^2  \tau \left( A_{(m+1)(m+1)} + A_{(m-1)(m-1)} - 2A_{mm} \right) + \epsilon^2 \left( {\rm convergent~term} \right).
\label{sec_reno_perturbed_D3}
\end{eqnarray}
To eliminate the divergent term, we introduce an arbitrary parameter $\tau'$ and replace a constant  $A_{mm}$ with a time-dependent variable $ A_{mm}(\tau')$. Then, Eq.~\eqref{sec_reno_perturbed_D3} is transformed into the following equation for a new perturbed solution $\bar{D}_{mm}'(\tau,\tau')$: 
\begin{eqnarray}
\bar{D}_{mm}'(\tau,\tau') = &&  A_{mm}(\tau')  +  i \epsilon \left( 1 - {\rm e}^{-\tau} \right)   \left( A_{m(m+1)} + A_{m(m-1)} - A_{(m+1)m} - A_{(m-1)m} \right) \nonumber \\
&&+ 2 \epsilon^2  (\tau - \tau') \left( A_{(m+1)(m+1)}(\tau') + A_{(m-1)(m-1)}(\tau') - 2A_{mm}(\tau') \right) + \epsilon^2 \left( {\rm convergent~term} \right).
\label{sec_reno_perturbed_D4}
\end{eqnarray}
Here, $\tau-\tau'$ in the third term on the right-hand side of Eq.~\eqref{sec_reno_perturbed_D4} plays a role in eliminating the divergence.
We require that Eq.~\eqref{sec_reno_perturbed_D4} is invariant under the choice of the parameter $\tau'$, i.e.,
\begin{eqnarray}
\frac{d}{d \tau'} \bar{D}_{mm}'(\tau,\tau') |_{\tau'=\tau} = 0. 
\label{sec_reno_perturbed_D5}
\end{eqnarray}
As a result, we finally obtain a second-order renormalized equation given by
\begin{eqnarray}
\frac{d}{d \tau} A_{mm}(\tau) = 2 \epsilon^2 \left( A_{(m+1)(m+1)}(\tau) + A_{(m-1)(m-1)}(\tau) - 2A_{mm}(\tau) \right) + \mathcal{O}(\epsilon^3).  
\label{sec_reno_perturbed_D6}
\end{eqnarray}
This means that the dynamics for the diagonal part of $D_{mn}(t)$ is well described by
\begin{eqnarray}
\frac{d}{d t} D_{mm}(t) \simeq \frac{2}{\gamma} \left( D_{(m+1)(m+1)}(t) + D_{(m-1)(m-1)}(t) - 2D_{mm}(t) \right). 
\label{sec_reno_perturbed_D6}
\end{eqnarray}

Next, we derive the effective equation for $F_{mnmn}(t)$. As with the above procedure, we rewrite Eq.~\eqref{f_F} with $\tau$ and $\bar{F}_{mnmn}(\tau) = F_{mnmn}(t)$:
\begin{eqnarray}
\frac{d}{d \tau} \bar{F}_{mnpq}(\tau) =~  \Lambda_{mnpq} \bar{F}_{mnpq}(\tau) +  i \epsilon & \Bigl( &  \bar{F}_{mn(p+1)q}(\tau) + \bar{F}_{mn(p-1)q}(\tau) + \bar{F}_{mnp(q+1)}(\tau) + \bar{F}_{mnp(q-1)}(\tau) \nonumber \\ 
 &-& \bar{F}_{(m+1)npq}(\tau) - \bar{F}_{(m-1)npq}(\tau) - \bar{F}_{m(n+1)pq}(\tau) - \bar{F}_{m(n-1)pq}(\tau)  \Bigl)
\label{sec_reno_perturbed_F1}
\end{eqnarray}
with $ \Lambda_{mnpq} =  \delta_{mq} + \delta_{mp}  + \delta_{nq} + \delta_{np}  - 2 $. 
The four-point correlation matrix is expanded as
\begin{eqnarray}
\bar{F}_{mnpq}(\tau) = \bar{F}_{mnpq}^{(0)}(\tau) + \epsilon \bar{F}_{mnpq}^{(1)}(\tau) + \epsilon^2 \bar{F}_{mnpq}^{(2)}(\tau) + \cdots, 
\label{sec_reno_perturbed_F2}
\end{eqnarray}
where $\bar{F}_{mnpq}^{(j)}(\tau)~(j=0,1,2,\cdots)$ is a $j$th-order four-point correlation matrix.
We substitute Eq.~\eqref{sec_reno_perturbed_F2} into Eq.~\eqref{sec_reno_perturbed_F1}, obtaining the following equation:
\begin{eqnarray}
&&\epsilon^0:~~\frac{d}{d \tau} \bar{F}_{mnpq}^{(0)}(\tau) = \Lambda_{mnpq} \bar{F}_{mnpq}^{(0)}(\tau),  \label{sec_reno_perturbed_F3}  \\
\nonumber \\
&&\epsilon^1:~~\frac{d}{d \tau} \bar{F}_{mnpq}^{(1)}(\tau) = \Lambda_{mnpq} \bar{F}_{mnpq}^{(1)}(\tau)  + i  \Bigl(   \bar{F}_{mn(p+1)q}^{(0)}(\tau) + \bar{F}_{mn(p-1)q}^{(0)}(\tau) + \bar{F}_{mnp(q+1)}^{(0)}(\tau) + \bar{F}_{mnp(q-1)}^{(0)}(\tau) \nonumber \\ 
\nonumber \\
&& \hspace{6.2cm} - \bar{F}_{(m+1)npq}^{(0)}(\tau) - \bar{F}_{(m-1)npq}^{(0)}(\tau) - \bar{F}_{m(n+1)pq}^{(0)}(\tau) - \bar{F}_{m(n-1)pq}^{(0)}(\tau)  \Bigl),  \label{sec_reno_perturbed_F4}  \\
\nonumber \\
&&\epsilon^2:~~\frac{d}{d \tau} \bar{F}_{mnpq}^{(2)}(\tau) = \Lambda_{mnpq} \bar{F}_{mnpq}^{(2)}(\tau)  + i  \Bigl(   \bar{F}_{mn(p+1)q}^{(1)}(\tau) + \bar{F}_{mn(p-1)q}^{(1)}(\tau) + \bar{F}_{mnp(q+1)}^{(1)}(\tau) + \bar{F}_{mnp(q-1)}^{(1)}(\tau) \nonumber \\ 
\nonumber \\
&& \hspace{6.2cm} - \bar{F}_{(m+1)npq}^{(1)}(\tau) - \bar{F}_{(m-1)npq}^{(1)}(\tau) - \bar{F}_{m(n+1)pq}^{(1)}(\tau) - \bar{F}_{m(n-1)pq}^{(1)}(\tau)  \Bigl),  \label{sec_reno_perturbed_F5}  \\
&& ~\cdot \nonumber \\
&& ~\cdot \nonumber \\
&& ~\cdot \nonumber \\
&&\epsilon^a:~~\frac{d}{d \tau} \bar{F}_{mnpq}^{(a)}(\tau) = \Lambda_{mnpq} \bar{F}_{mnpq}^{(a)}(\tau)  + i  \Bigl(   \bar{F}_{mn(p+1)q}^{(a-1)}(\tau) + \bar{F}_{mn(p-1)q}^{(a-1)}(\tau) + \bar{F}_{mnp(q+1)}^{(a-1)}(\tau) + \bar{F}_{mnp(q-1)}^{(a-1)}(\tau) \nonumber \\ 
&& \hspace{6.2cm} - \bar{F}_{(m+1)npq}^{(a-1)}(\tau) - \bar{F}_{(m-1)npq}^{(a-1)}(\tau) - \bar{F}_{m(n+1)pq}^{(a-1)}(\tau) - \bar{F}_{m(n-1)pq}^{(a-1)}(\tau)  \Bigl).  \label{sec_reno_perturbed_F6} 
\end{eqnarray}
We solve Eqs.~\eqref{sec_reno_perturbed_F3},~\eqref{sec_reno_perturbed_F4}, and~\eqref{sec_reno_perturbed_F5} to obtain
\begin{eqnarray}
&& \bar{F}_{mnpq}^{(0)}(\tau) = B_{mnpq} {\rm e}^{ \Lambda_{mnpq} \tau },  \label{sec_reno_perturbed_F7} \\
\nonumber \\
&& \bar{F}_{mnpq}^{(1)}(\tau) =   i  {\rm e}^{ \Lambda_{mnpq} \tau } \int_{0}^{\tau} dT {\rm e}^{ - \Lambda_{mnpq} T } \Bigl( \bar{F}_{mnp(q+1)}^{(0)}(T) + \bar{F}_{mnp(q-1)}^{(0)}(T) + \bar{F}_{mn(p+1)q}^{(0)}(T) + \bar{F}_{mn(p-1)q}^{(0)}(T)  \nonumber \\
\nonumber \\
&& \hspace{6.0cm}- \bar{F}_{m(n+1)pq}^{(0)}(T) - \bar{F}_{m(n-1)pq}^{(0)}(T) - \bar{F}_{(m+1)npq}^{(0)}(T) - \bar{F}_{(m-1)npq}^{(0)}(T)
\Bigl),  \label{sec_reno_perturbed_F8} \\
\nonumber \\
&& \bar{F}_{mnpq}^{(2)}(\tau) =   i  {\rm e}^{ \Lambda_{mnpq} \tau } \int_{0}^{\tau} dT {\rm e}^{ - \Lambda_{mnpq} T } \Bigl( \bar{F}_{mnp(q+1)}^{(1)}(T) + \bar{F}_{mnp(q-1)}^{(1)}(T) + \bar{F}_{mn(p+1)q}^{(1)}(T) + \bar{F}_{mn(p-1)q}^{(1)}(T)  \nonumber \\
&& \hspace{6.0cm}- \bar{F}_{m(n+1)pq}^{(1)}(T) - \bar{F}_{m(n-1)pq}^{(1)}(T) - \bar{F}_{(m+1)npq}^{(1)}(T) - \bar{F}_{(m-1)npq}^{(1)}(T)
\Bigl).  \label{sec_reno_perturbed_F9} 
\end{eqnarray}
with a constant $B_{mnpq}$. Then, for $|n-m|>2$, we obtain the second-order perturbed equation:
\begin{eqnarray}
\bar{F}_{mnmn}(\tau) = &&  B_{mnmn}  + 2 \epsilon^2  \tau \left( B_{(m+1)n(m+1)n} + B_{(m-1)n(m-1)n} + B_{m(n+1)m(n+1)} + B_{m(n-1)m(n-1)} - 4B_{mnmn} \right) \nonumber \\
&& + \epsilon \left( {\rm convergent~term} \right) + \epsilon^2 \left( {\rm convergent~term} \right), 
\label{sec_reno_perturbed_F10}
\end{eqnarray}
from which one can see that the second term on the right-hand side is divergent for $\tau \rightarrow \infty$. 
In order to eliminate the divergence, we introduce an arbitrary parameter $\tau'$ and replace a constant $B_{mnmn}$ with a time-dependent variable $B_{mnmn}(\tau')$, and Eq.~\eqref{sec_reno_perturbed_F10} becomes an equation for a new variable $\bar{F}_{mnmn}'(\tau,\tau')$:
\begin{eqnarray}
\bar{F}_{mnmn}'(\tau,\tau') = &&  B_{mnmn}(\tau')  + 2 \epsilon^2  (\tau - \tau')\Bigl( B_{(m+1)n(m+1)n}(\tau') + B_{(m-1)n(m-1)n}(\tau') \nonumber \\
&& + B_{m(n+1)m(n+1)}(\tau') + B_{m(n-1)m(n-1)}(\tau') - 4B_{mnmn}(\tau') \Bigl) \nonumber \\
&& + \epsilon \left( {\rm convergent~term} \right) + \epsilon^2 \left( {\rm convergent~term} \right).
\label{sec_reno_perturbed_F11}
\end{eqnarray}
We require that Eq.~\eqref{sec_reno_perturbed_F11} is independent of $\tau'$, obtaining
\begin{eqnarray}
\frac{d}{d \tau'} \bar{F}_{mnmn}'(\tau,\tau') |_{\tau'=\tau} = 0. 
\label{sec_reno_perturbed_F12}
\end{eqnarray}
Therefore, we derive the second-order renormalized equation given by
\begin{eqnarray}
&& \frac{d}{d \tau} B_{mnmn}(\tau)  = 2 \epsilon^2  \Bigl( B_{(m+1)n(m+1)n}(\tau) + B_{(m-1)n(m-1)n}(\tau) \nonumber \\
&& \hspace{3.1cm} + B_{m(n+1)m(n+1)}(\tau) + B_{m(n-1)m(n-1)}(\tau) - 4B_{mnmn}(\tau) \Bigl) + \mathcal{O}(\epsilon^3).
\label{sec_reno_perturbed_F13}
\end{eqnarray}
Thus, the effective equation for $F_{mnmn}(t) $ becomes
\begin{eqnarray}
\frac{d}{d t} F_{mnmn}(t)  \simeq \frac{2}{\gamma}  \Bigl( F_{(m+1)n(m+1)n}(t) + F_{(m-1)n(m-1)n}(t) + F_{m(n+1)m(n+1)}(t) + F_{m(n-1)m(n-1)}(t) - 4F_{mnmn}(t) \Bigl). \nonumber \\
\label{sec_reno_perturbed_F14}
\end{eqnarray}
Finally, we comment on the effective equation for $|m-n| \leq 2$. Currently, we do not succeed in deriving the equation, and we expect that our method needs to be modified for the short-scale dynamics. As numerically discussed later, the short-scale dynamics does not affect the scaling property of the surface roughness. 

\subsubsection{bosonic system}
We apply the same renormalization-group method to Eqs.~\eqref{b_D} and \eqref{b_F}, obtaining the same effective equations for $|m-n|>2$:
\begin{eqnarray}
\frac{d}{d t} D_{mm}(t) \simeq \frac{2}{\gamma} \left( D_{(m+1)(m+1)}(t) + D_{(m-1)(m-1)}(t) - 2D_{mm}(t) \right), 
\label{sec_reno_perturbed_D_boson}
\end{eqnarray}
\begin{eqnarray}
\frac{d}{d t} F_{mnmn}(t)  \simeq \frac{2}{\gamma}  \Bigl( F_{(m+1)n(m+1)n}(t) + F_{(m-1)n(m-1)n}(t) + F_{m(n+1)m(n+1)}(t) + F_{m(n-1)m(n-1)}(t) - 4F_{mnmn}(t) \Bigl). \nonumber \\
\label{sec_reno_perturbed_F_boson}
\end{eqnarray}

\subsubsection{numerical test}
\begin{figure}[t]
\begin{center}
\includegraphics[keepaspectratio, width=17.5cm,clip]{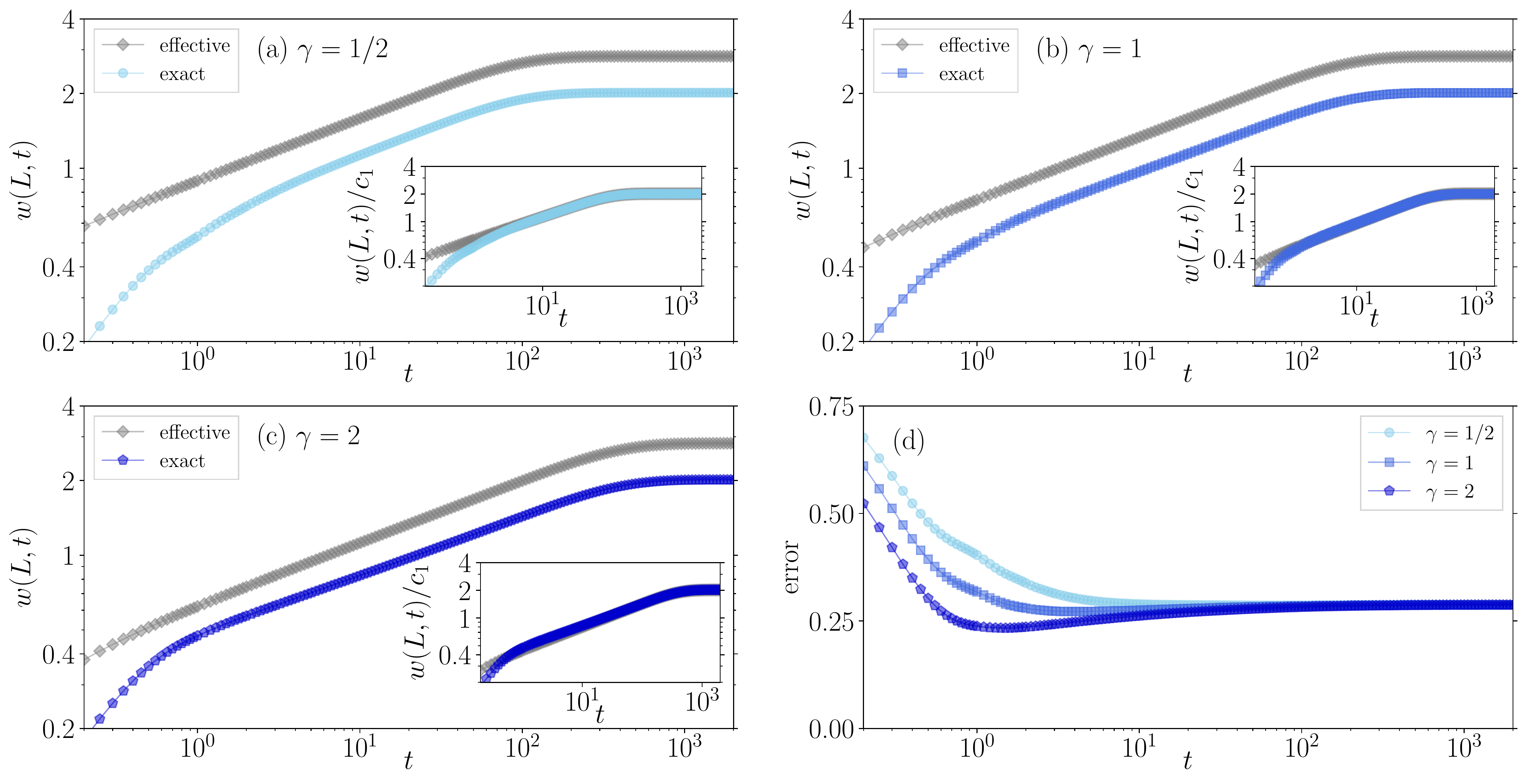}
\caption{Exact and effective dynamics of the surface roughness in the fermions with the dephasing. 
The system size is set to be $L=64$, and the strength of the dephasing is (a) $\gamma=0.5$, (b) $\gamma=1$, and (c) $\gamma=2$.
The initial state is the staggered state. The insets of (a), (b), and (c) show the graphs where the ordinates are multiplied by a constant $c_1$ such that the saturated roughness becomes almost equal. In (d), we display the relative error between $w(L,t)$ and $w_{\rm eff}(L,t)$ defined in Eq.~\eqref{error}. 
} 
\label{figureS1} 
\end{center}
\end{figure}

\begin{figure}[t]
\begin{center}
\includegraphics[keepaspectratio, width=17.5cm,clip]{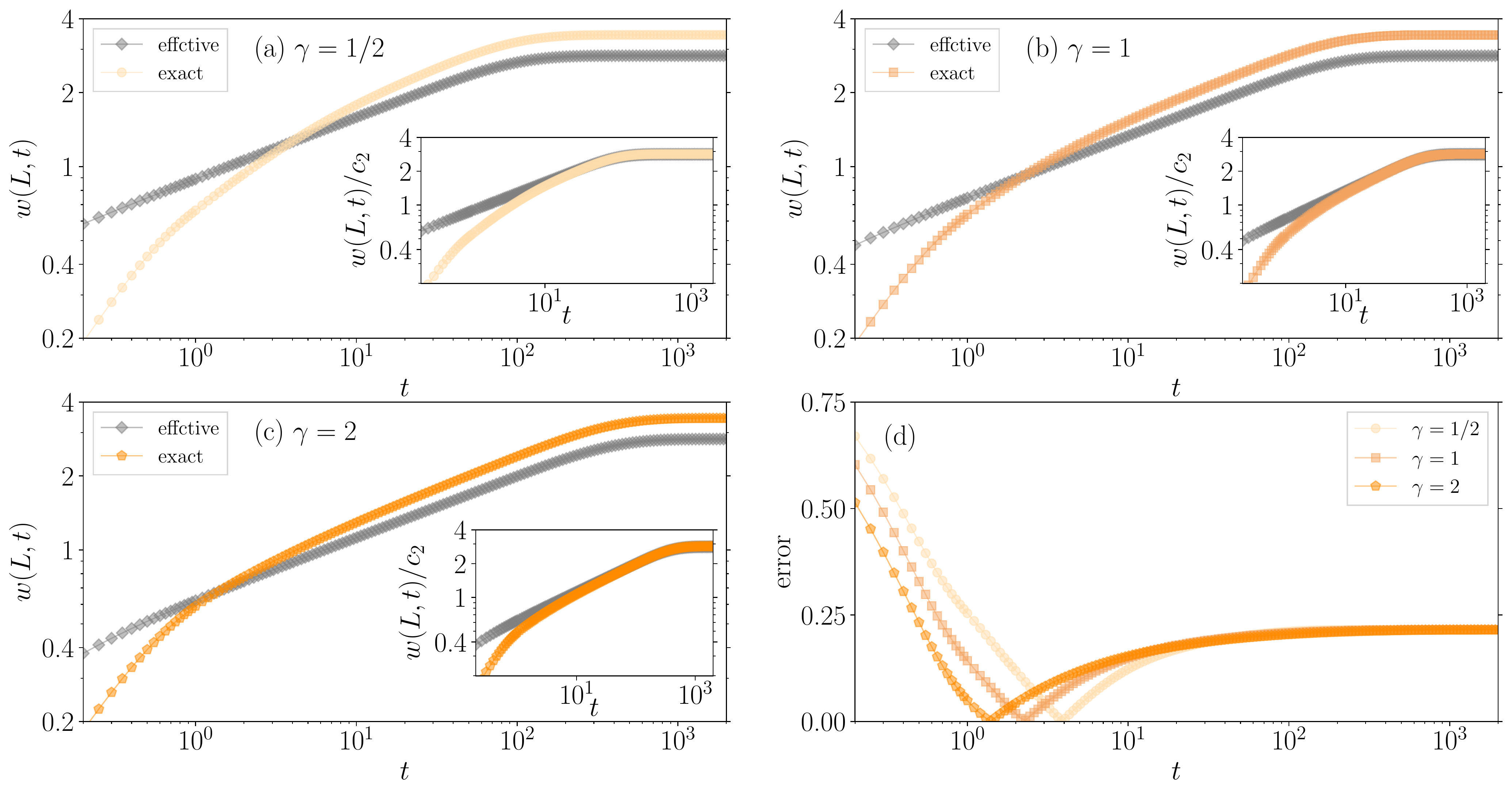}
\caption{
Exact and effective dynamics of the surface roughness in the bosons with the dephasing. 
The system size is set to be $L=64$, and the strength of the dephasing is (a) $\gamma=0.5$, (b) $\gamma=1$, and (c) $\gamma=2$. 
The insets of (a), (b), and (c) show the graphs where the ordinates are multiplied by a constant $c_2$ such that the saturated roughness becomes almost equal.
The initial state is the staggered state. In (d), we display the relative error between $w(L,t)$ and $w_{\rm eff}(L,t)$ defined in Eq.~\eqref{error}. 
} 
\label{figureS2} 
\end{center}
\end{figure}

We numerically test our effective equations of motion by comparing the surface roughness $w(L,t)$ and $w_{\rm eff}(L,t)$, which are obtained in the exact calculation and the effective one, respectively. In the numerical calculations for $F_{mnmn}(t)$, we use the effective equation even for $|m-n| \leq 2$. Figures~\ref{figureS1}(a-c) and \ref{figureS2} (a-c) exhibit the time evolutions of $w(L,t)$ and $w_{\rm eff}(L,t)$ for the fermions and the bosons starting from the staggered state. We find that the effective surface-roughness well obeys the  power-law growth with $t^{0.25}$ and that the scaling functions of $w_{\rm eff}(L,t)$ are almost identical to those of $w(L,t)$ as shown in the insets, where we multiply the ordinates of Fig.~\ref{figureS1}(a-c) and \ref{figureS2}(a-c) by constants $c_1$ and $c_2$ in order that the effective surface roughness $w_{\rm eff}(L,t)$ is collapsed to the exact one $w(L,t)$. To evaluate the deviation of $w(L,t)$ from $w_{\rm eff}(L,t)$, we define the relative error
\begin{eqnarray}
\frac{|w(L,t)-w_{\rm eff}(L,t)|}{w_{\rm eff}(L,t)}. 
\label{error}
\end{eqnarray}
Figures~\ref{figureS1}(d) and \ref{figureS2}(d) show the relative errors, which are smaller than $0.3$ in $ 4 \lesssim t$. Note that the surface roughness grows with $t^{1/4}$ in this timescale.

We consider that this deviation comes from the fact that our effective equations cannot correctly describe the short-range dynamics because the effective equations of motion for $F_{mnmn}(t)$ are valid for $|m-n|>2$. However, the scaling properties are not affected by this short-range physics at all, and our effective equations can reproduce the proper scaling exponents and the scaling function as shown in the insets of Figs.~\ref{figureS1} and \ref{figureS2}. 

\clearpage

\section{Derivation of the effective diffusive equations for the Lindblad equation without the strong-dephasing condition}
We derive effective equations of motion for the two- and four-point correlation matrices without relying on the strong dephasing $\gamma \gg 1$. 
The assumptions used here are the following:
\begin{enumerate}
  \item[ ] (A1) $ t\gamma \gg 1 $, 
  \item[ ] (A2) $ |D_{mm}(t)| \gg |D_{pq}(t)| $ for $p\neq q$,
  \item[ ] (A3) $ |D_{mm}(t)| \gg | e^{-\gamma t} \int_0^{t} dT e^{-\gamma \lambda_{pq} T} D_{pq}(T) | $ for $p \neq q$, 
  \item[ ] (A4) $ |F_{mnmn}(t)| \gg |F_{pqrs}(t)| $ for $(p,q,r,s)$ not satisfying $p = r$ and $q = s$, 
  \item[ ] (A5) $ |F_{mnmn}(t)| \gg |  e^{-\gamma t} \int_0^{t} dT e^{-\gamma \Lambda_{pqrs} T} F_{pqrs}(T) | $ for $(p,q,r,s)$ not satisfying $p = r$ and $q = s$.
\end{enumerate}
These assumptions are plausible in the dephasing Lindblad equation because the off-diagonal elements of the correlation matrices decay exponentially due to the loss of coherence while the diagonal elements $D_{mm}(t)$ and $F_{mnmn}(t)$ do not.  

\subsubsection{fermionic system}
We derive an effective equation using the above five assumptions, Eqs.~\eqref{f_D}, and \eqref{f_F}. First, using the equation of motion for $D_{mn}$, we analytically obtain the effective equation for the diagonal part $D_{mm}$. Second, we apply the same procedure to the equation of motion for $F_{mnpq}$, deriving the effective equation for $F_{mnmn}$. 

We introduce $D_{mn} = \tilde{D}_{mn} {\rm e}^{ \gamma \lambda_{mn} t}$ to obtain 
\begin{eqnarray}
\frac{d}{d t} \tilde{D}_{mn} = i {\rm e}^{-\gamma \lambda_{mn} t} \left( {\rm e}^{\gamma \lambda_{m(n+1)} t} \tilde{D}_{m(n+1)} + {\rm e}^{\gamma \lambda_{m(n-1)} t} \tilde{D}_{m(n-1)} - {\rm e}^{\gamma \lambda_{(m+1)n} t} \tilde{D}_{(m+1)n}  - {\rm e}^{\gamma \lambda_{(m-1)n} t} \tilde{D}_{(m-1)n}  \right).
\label{f_D1}
\end{eqnarray}
In order to derive the effective equation of motion for the diagonal part $D_{mm}$, we explicitly write Eq.~\eqref{f_D1} for $\tilde{D}_{mm}$, $\tilde{D}_{(m+1)m}$, $\tilde{D}_{(m-1)m}$, $\tilde{D}_{m(m+1)}$, and $\tilde{D}_{m(m-1)}$: 
\begin{eqnarray}
\frac{d}{d t} \tilde{D}_{mm} = i {\rm e}^{-\gamma t} \left(  \tilde{D}_{m(m+1)} + \tilde{D}_{m(m-1)} - \tilde{D}_{(m+1)m} - \tilde{D}_{(m-1)m}  \right), 
\label{f_D2}
\end{eqnarray}
\begin{eqnarray}
\frac{d}{d t} \tilde{D}_{m(m+1)} = i  \left(  {\rm e}^{\gamma t} \tilde{D}_{mm} - {\rm e}^{\gamma t} \tilde{D}_{(m+1)(m+1)} + \tilde{D}_{m(m+2)} - \tilde{D}_{(m-1)(m+1)}  \right), 
\label{f_D3}
\end{eqnarray}
\begin{eqnarray}
\frac{d}{d t} \tilde{D}_{m(m-1)} = i  \left(  {\rm e}^{\gamma t} \tilde{D}_{mm} - {\rm e}^{\gamma t} \tilde{D}_{(m-1)(m-1)} + \tilde{D}_{m(m-2)} - \tilde{D}_{(m+1)(m-1)}  \right), 
\label{f_D4}
\end{eqnarray}
\begin{eqnarray}
\frac{d}{d t} \tilde{D}_{(m+1)m} = i  \left(  {\rm e}^{\gamma t} \tilde{D}_{(m+1)(m+1)} - {\rm e}^{\gamma t} \tilde{D}_{mm} - \tilde{D}_{(m+2)m} + \tilde{D}_{(m+1)(m-1)}  \right), 
\label{f_D5}
\end{eqnarray}
\begin{eqnarray}
\frac{d}{d t} \tilde{D}_{(m-1)m} = i  \left(  {\rm e}^{\gamma t} \tilde{D}_{(m-1)(m-1)} - {\rm e}^{\gamma t} \tilde{D}_{mm} - \tilde{D}_{(m-2)m} + \tilde{D}_{(m-1)(m+1)}  \right). 
\label{f_D6}
\end{eqnarray}

Here, we evaluate ${\rm e}^{-\gamma t}  \tilde{D}_{m(m+1)}(t) $ in Eq.~\eqref{f_D2} by integrating Eq.~\eqref{f_D3} as 
\begin{eqnarray}
{\rm e}^{-\gamma t}  \tilde{D}_{m(m+1)} (t) &=&  i \int_{0}^{t} dt_1 {\rm e}^{-\gamma t}  \left(  {\rm e}^{\gamma t_1} \tilde{D}_{mm}(t_1) - {\rm e}^{\gamma t_1} \tilde{D}_{(m+1)(m+1)}(t_1) + \tilde{D}_{m(m+2)}(t_1) - \tilde{D}_{(m-1)(m+1)}(t_1)  \right) \nonumber \\
&\simeq& i \int_{0}^{t} dt_1 {\rm e}^{\gamma (t_1-t)} \left(   \tilde{D}_{mm}(t_1) -  \tilde{D}_{(m+1)(m+1)}(t_1) \right)  \nonumber \\
&\simeq& \frac{i}{\gamma} \left(   \tilde{D}_{mm}(t) -  \tilde{D}_{(m+1)(m+1)}(t) \right) - \frac{i}{\gamma} \int_{0}^{t}  dt_1 {\rm e}^{\gamma (t_1-t)} \frac{d}{dt_1}\left(   \tilde{D}_{mm}(t_1) -  \tilde{D}_{(m+1)(m+1)}(t_1) \right) \nonumber \\
&=& \frac{i}{\gamma} \left(   \tilde{D}_{mm}(t) -  \tilde{D}_{(m+1)(m+1)}(t) \right) + \frac{ {\rm e}^{-\gamma t}   }{\gamma} \int_{0}^{t}  dt_1 \Bigl(   \tilde{D}_{m(m+1)}(t_1) + \tilde{D}_{m(m-1)}(t_1) - \tilde{D}_{(m+1)m}(t_1)   \nonumber \\
&& - \tilde{D}_{(m-1)m}(t_1) - \tilde{D}_{(m+1)(m+2)}(t_1) - \tilde{D}_{(m+1)m}(t_1) + \tilde{D}_{(m+2)(m+1)}(t_1) + \tilde{D}_{m(m+1)}(t_1) \Bigl)  \nonumber \\
&\simeq& \frac{i}{\gamma} \left(   \tilde{D}_{mm}(t) -  \tilde{D}_{(m+1)(m+1)}(t) \right).
\label{f_D7}
\end{eqnarray}
Here, we use the assumption (A2) in the first line, the assumption (A1) in the second line, Eq.~\eqref{f_D2} in the third line, and the assumption (A3) in the fourth and fifth lines. Following the same way, we obtain the approximate expressions:
\begin{eqnarray}
{\rm e}^{-\gamma t}  \tilde{D}_{m(m-1)} (t) \simeq \frac{i}{\gamma} \left(   \tilde{D}_{mm}(t) -  \tilde{D}_{(m-1)(m-1)}(t) \right), 
\label{f_D8}
\end{eqnarray}
\begin{eqnarray}
{\rm e}^{-\gamma t}  \tilde{D}_{(m+1)m} (t) \simeq \frac{i}{\gamma} \left(   \tilde{D}_{(m+1)(m+1)}(t) -  \tilde{D}_{mm}(t) \right), 
\label{f_D9}
\end{eqnarray}
\begin{eqnarray}
{\rm e}^{-\gamma t}  \tilde{D}_{(m-1)m} (t) \simeq \frac{i}{\gamma} \left(   \tilde{D}_{(m-1)(m-1)}(t) -  \tilde{D}_{mm}(t) \right).
\label{f_D10}
\end{eqnarray}
Substituting Eqs.~\eqref{f_D7}--\eqref{f_D10} into Eq.~\eqref{f_D2}, we obtain the effective equation:
\begin{eqnarray}
\frac{d}{dt} D_{mm}(t)  \simeq \frac{2}{\gamma} \left(  {D}_{(m+1)(m+1)}(t) + {D}_{(m-1)(m-1)}(t)  -  2 {D}_{mm}(t) \right). 
\label{f_D11}
\end{eqnarray}
This is identical to the effective equation~\eqref{sec_reno_perturbed_D6} obtained via the renormalization-group method. 

Next, we derive an effective equation of motion for $F_{mnmn}$. 
In what follows, we assume $|m-n|>2$ to derive the effective equation. 

Using the transformation $F_{mnpq} = \tilde{F}_{mnpq} {\rm e}^{\gamma \Lambda_{mnpq} t}$, we obtain the equation for $\tilde{F}_{mnpq}$ from Eq.~\eqref{f_F}:
\begin{eqnarray}
\frac{d}{d t} \tilde{F}_{mnpq} = i  {\rm e}^{-\gamma \Lambda_{mnpq} t} (&& {\rm e}^{\gamma \Lambda_{mn(p+1)q} t}  \tilde{F}_{mn(p+1)q} + {\rm e}^{\gamma \Lambda_{mn(p-1)q} t} \tilde{F}_{mn(p-1)q} \nonumber \\
&& + {\rm e}^{\gamma \Lambda_{mnp(q+1)} t} \tilde{F}_{mnp(q+1)} + {\rm e}^{\gamma \Lambda_{mnp(q-1)} t} \tilde{F}_{mnp(q-1)}  \nonumber \\
&& - {\rm e}^{\gamma \Lambda_{(m+1)npq} t} \tilde{F}_{(m+1)npq} - {\rm e}^{\gamma \Lambda_{(m-1)npq} t} \tilde{F}_{(m-1)npq} \nonumber \\
&& - {\rm e}^{\gamma \Lambda_{m(n+1)pq} t} \tilde{F}_{m(n+1)pq} - {\rm e}^{\gamma \Lambda_{m(n-1)pq} t} \tilde{F}_{m(n-1)pq} ).
\label{f_F1}
\end{eqnarray}
Then, the equation for $\tilde{F}_{mnmn}$ for $|m-n| > 2$ reads
\begin{eqnarray}
\frac{d}{d t} \tilde{F}_{mnmn} = i  {\rm e}^{- \gamma t} (  &&  \tilde{F}_{mnm(n+1)}  + \tilde{F}_{mnm(n-1)} + \tilde{F}_{mn(m+1)n} +  \tilde{F}_{mn(m-1)n}  \nonumber \\
&& -  \tilde{F}_{m(n+1)mn} -  \tilde{F}_{m(n-1)mn} - \tilde{F}_{(m+1)nmn} -  \tilde{F}_{(m-1)nmn} ).
\label{f_F2}
\end{eqnarray}

We first evaluate the term ${\rm e}^{- \gamma t} \tilde{F}_{mnm(n+1)}$ on the right-hand side of Eq.~\eqref{f_F2}. The explicit equation of motion for $\tilde{F}_{mnm(n+1)}$ reads
\begin{eqnarray}
\frac{d}{d t} \tilde{F}_{mnm(n+1)} = i  ( && {\rm e}^{\gamma t} \tilde{F}_{mnmn} - {\rm e}^{\gamma t} \tilde{F}_{m(n+1)m(n+1)} + \tilde{F}_{mnm(n+2)} - \tilde{F}_{m(n-1)m(n+1)} + {\rm e}^{-\gamma t} \tilde{F}_{mn(m+1)(n+1)} \nonumber \\
 && + {\rm e}^{-\gamma t} \tilde{F}_{mn(m-1)(n+1)} - {\rm e}^{-\gamma t} \tilde{F}_{(m+1)nm(n+1)} - {\rm e}^{-\gamma t} \tilde{F}_{(m-1)nm(n+1)} )
\label{f_F3}
\end{eqnarray}
for $|m-n| > 2$. Then, using the assumptions (A1), (A4), and (A5) and following the same procedure shown in the case of the two-point correlation matrices, we get the following approximated expression:
\begin{eqnarray}
{\rm e}^{- \gamma t} \tilde{F}_{mnm(n+1)} \simeq \frac{i}{\gamma}  \left(   \tilde{F}_{mnmn}  -  \tilde{F}_{m(n+1)m(n+1)}  \right).
\label{f_F4}
\end{eqnarray}
In the same way, we obtain
\begin{eqnarray}
{\rm e}^{- \gamma t} \tilde{F}_{mnm(n-1)} \simeq \frac{i}{\gamma}  \left(   \tilde{F}_{mnmn}  -  \tilde{F}_{m(n-1)m(n-1)}  \right),
\label{f_F5}
\end{eqnarray}
\begin{eqnarray}
{\rm e}^{- \gamma t} \tilde{F}_{mn(m+1)n} \simeq \frac{i}{\gamma}  \left(   \tilde{F}_{mnmn}  -  \tilde{F}_{(m+1)n(m+1)n}  \right), 
\label{f_F6}
\end{eqnarray}
\begin{eqnarray}
{\rm e}^{- \gamma t} \tilde{F}_{mn(m-1)n} \simeq \frac{i}{\gamma}  \left(   \tilde{F}_{mnmn}  -  \tilde{F}_{(m-1)n(m-1)n}  \right), 
\label{f_F7}
\end{eqnarray}
\begin{eqnarray}
{\rm e}^{- \gamma t} \tilde{F}_{(m+1)nmn} \simeq \frac{i}{\gamma}  \left(   \tilde{F}_{(m+1)n(m+1)n} - \tilde{F}_{mnmn}  \right),
\label{f_F8}
\end{eqnarray}
\begin{eqnarray}
{\rm e}^{- \gamma t} \tilde{F}_{(m-1)nmn} \simeq \frac{i}{\gamma}  \left(   \tilde{F}_{(m-1)n(m-1)n} - \tilde{F}_{mnmn}  \right), 
\label{f_F9}
\end{eqnarray}
\begin{eqnarray}
{\rm e}^{- \gamma t} \tilde{F}_{m(n+1)mn} \simeq \frac{i}{\gamma}  \left(   \tilde{F}_{m(n+1)m(n+1)} - \tilde{F}_{mnmn}  \right), 
\label{f_F10}
\end{eqnarray}
\begin{eqnarray}
{\rm e}^{- \gamma t} \tilde{F}_{m(n-1)mn} \simeq \frac{i}{\gamma}  \left(   \tilde{F}_{m(n-1)m(n-1)} - \tilde{F}_{mnmn}  \right).
\label{f_F10}
\end{eqnarray}

Substituting Eqs.~\eqref{f_F4}--\eqref{f_F10} into Eq.~\eqref{f_F2}, we finally obtain the effective equation of motion. 
\begin{eqnarray}
\frac{d}{dt} F_{mnmn}  \simeq \frac{2}{\gamma} \left( F_{(m+1)n(m+1)n}  + F_{(m-1)n(m-1)n} + F_{m(n+1)m(n+1)} + F_{m(n-1)m(n-1)}  -  4 F_{mnmn} \right). 
\label{f_F11}
\end{eqnarray}
This is identical to the effective equation~\eqref{sec_reno_perturbed_F14} obtained via the renomarlization-group method. 

\subsubsection{bosonic system}
Using the same method discussed above, we find that the effective equations for $D_{mm}$ and $F_{mnmn}$ with $|m-n|>2$ are the same as the fermionic equation:
\begin{eqnarray}
\frac{d}{dt} D_{mm}  \simeq \frac{2}{\gamma} \left(  {D}_{(m+1)(m+1)} + {D}_{(m-1)(m-1)}  -  2 {D}_{mm} \right), 
\end{eqnarray}
\begin{eqnarray}
\frac{d}{dt} F_{mnmn}  \simeq \frac{2}{\gamma} \left( F_{(m+1)n(m+1)n}  + F_{(m-1)n(m-1)n} + F_{m(n+1)m(n+1)} + F_{m(n-1)m(n-1)}  -  4 F_{mnmn} \right). 
\end{eqnarray}

\subsubsection{Numerical test for the assumptions (A2)-(A5)}
We numerically investigate validity of the assumptions (A2)-(A5) used above. 
For this purpose, we first define the following quantities:
\begin{eqnarray}
G_{mn}(t) = e^{-\gamma t} \int_0^{t} dT e^{-\gamma \lambda_{mn} T} D_{mn}(T), 
\end{eqnarray}
\begin{eqnarray}
I_{mnpq}(t) = e^{-\gamma t} \int_0^{t} dT e^{-\gamma \Lambda_{mnpq} T} F_{mnpq}(T).
\end{eqnarray}
Differentiating them with respect to time $t$, we obatin
\begin{eqnarray}
\frac{d}{dt}G_{mn}(t) = -\gamma G_{mn}(t) + e^{-\gamma(\lambda_{mn}+1) t} D_{mn}(t), 
\end{eqnarray}
\begin{eqnarray}
\frac{d}{dt}I_{mnpq}(t) = -\gamma I_{mnpq}(t) + e^{-\gamma(\Lambda_{mnpq}+1) t} F_{mnpq}(t).
\end{eqnarray}
Solving these equations of motion with the initial conditions $G_{mn}(0)=0$ and $F_{mnpq}(0)=0$, we numerically evaluate $G_{mn}(t)$ and $I_{mnpq}(t)$, from which we can study how valid the assumptions (A3) and (A5) are.

\begin{figure}[t]
\begin{center}
\includegraphics[keepaspectratio, width=18cm,clip]{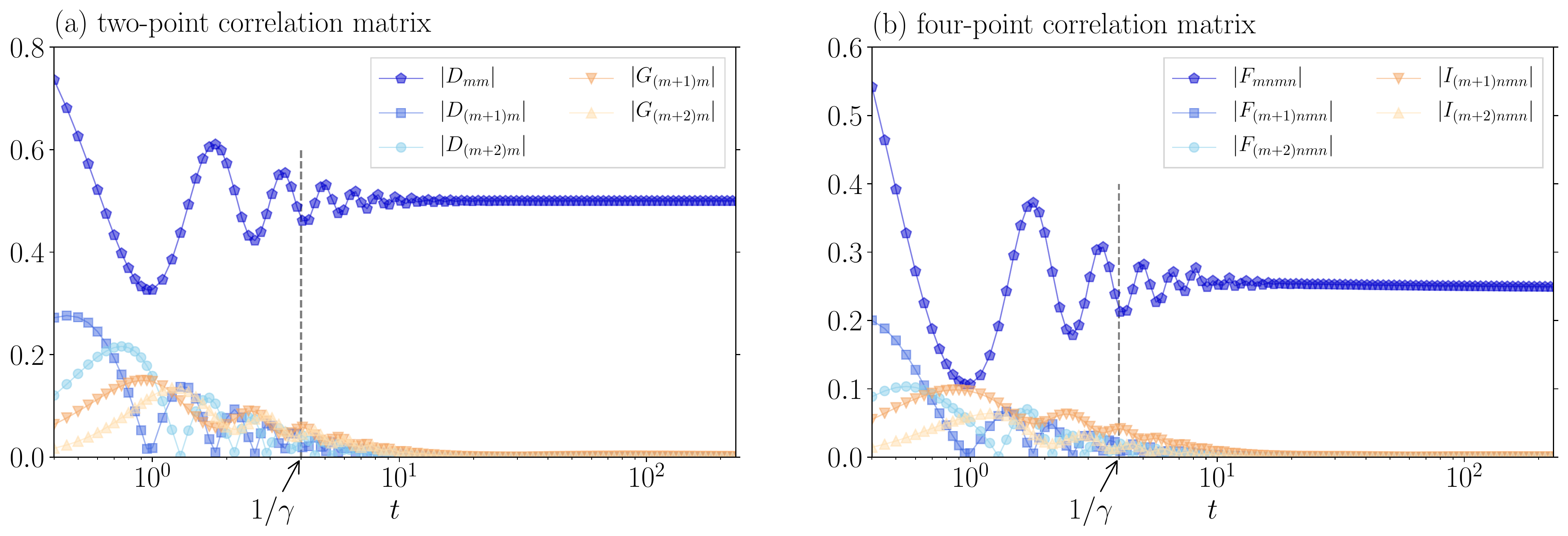}
\caption{Time evolutions of (a) $|D_{mm}(t)|$, $|D_{(m+1)m}(t)|$, $|D_{(m+2)m}(t)|$, $|G_{(m+1)m}(t)|$, $|G_{(m+2)m}(t)|$, and (b) $|F_{mnmn}(t)|$, $|F_{(m+1)nmn}(t)|$, $|F_{(m+2)nmn}(t)|$, $|I_{(m+1)nmn}(t)|$, $|I_{(m+2)nmn}(t)|$ in the dephasing fermionic dynamics starting from the staggered state. The parameters are $L=128$ and $\gamma=1/4$, which correspond to the lightest color of Fig.~3(a) in the main text. The integers $m$ and $n$ used in (a) and (b) are $m=L/2$ and $n=L/2 + 50$. The dashed vertical lines denote the dephasing time scale $1/\gamma$.
} 
\label{assumption_check} 
\end{center}
\end{figure}

Figure~\ref{assumption_check} shows the time evolutions of $|D_{mn}(t)|$, $|G_{mn}(t)|$, $|F_{mnpq}(t)|$, and $|I_{mnpq}(t)|$ in the fermions with dephasing ($\gamma=1/4$).
The initial state is the staggered state, and the system size is $L=128$. We find that the elements $|D_{mm}(t)|$ and $|F_{mnmn}(t)|$, which are the relevant variables in the effective equations, are dominant compared with the others when the time $t$ is larger than the dephasing time scale $1/\gamma$. Thus, we numerically confirm that the assumptions (A2)-(A5) are valid in the case with $\gamma=1/4$.

\clearpage
\section{Surface-roughness dynamics only with the out-flow of particles}
\begin{figure}[t]
\begin{center}
\includegraphics[keepaspectratio, width=18cm,clip]{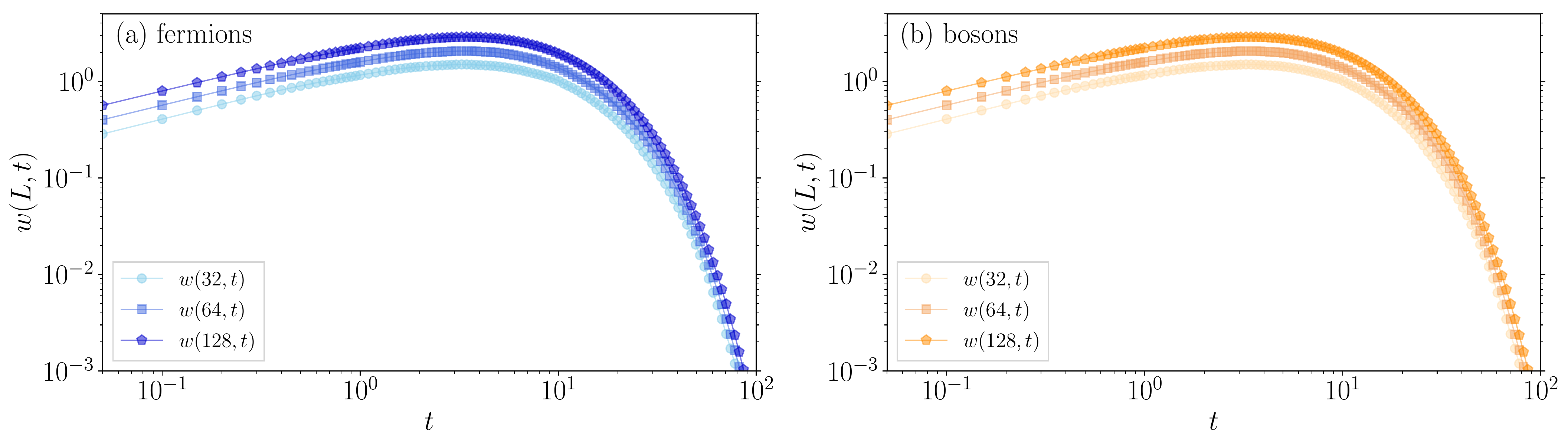}
\caption{Time evolutions of the surface roughness in (a) fermions and (b) bosons only with the out-flow of particles. The parameters are set to be $\gamma_{\rm in} = 0$ and $\gamma_{\rm out}=0.2$, and the initial state is the staggered state. 
} 
\label{particle_loss} 
\end{center}
\end{figure}

We show numerical results for the surface-roughness dynamics affected only by the out-flow of particles.
Figure~\ref{particle_loss} exhibits the time evolution of the surface roughness for the bosons and the fermons with $\gamma_{\rm in}=0$ and $\gamma_{\rm out}=0.2$. In this case, the total particle number decreases in time, and correspondingly the surface roughness finally approaches zero. 
Similar to Fig.~4 in the main text, we cannot find the conventional FV scaling. 

\end{document}